\documentclass[
10pt,nofootinbib,
aps,prd
,showpacs,
onecolumn,
]{revtex4-1}

\usepackage{hyperref}
\usepackage{verbatim}
\usepackage{amsmath}
\usepackage{latexsym}
\usepackage{revsymb}
\usepackage{yfonts}
\usepackage{soul}
\usepackage{natbib}
\usepackage{amsfonts}
\usepackage{amsmath}
\usepackage{amssymb}
\usepackage{amsthm}
\usepackage{graphicx}
\usepackage{bm}
\usepackage{bbm}

\def\beq{\begin{eqnarray}}
\def\eeq{\end{eqnarray}}

\newcommand{\be}{\begin{eqnarray} \begin{aligned}}
\newcommand{\ee}{\end{aligned} \end{eqnarray} }

\usepackage{amsfonts}

\begin{document}

\title{A perturbative approach to the spectral zeta functions of strings, 
drums and quantum billiards}
\author{Paolo Amore}
\email{paolo.amore@gmail.com}
\affiliation{Facultad de Ciencias, CUICBAS, Universidad de Colima, \\
Bernal D\'{\i}az del Castillo 340, Colima, Colima, Mexico} 

\begin{abstract}
We show that the spectral zeta functions of inhomogeneous strings and drums can
be calculated using Rayleigh-Schr\"odinger perturbation theory.
The inhomogeneities that can be treated with this method are small but otherwise
arbitrary and include the previously studied case of a piecewise constant density. 
In two dimensions the method can be used to derive the spectral zeta function of 
a domain obtained from the small deformation of a square. We also obtain exact 
sum rules that are valid for arbitrary densities and that correspond to the values 
taken by the spectral zeta function at integer positive values; we have tested 
numerically these sum rules in specific examples.
We  show that the Dirichlet or Neumann Casimir energies of an inhomogeneous string, evaluated to 
first order in perturbation theory, contain in some cases an irremovable divergence, but that the combination of the two is always free of divergences.
Finally, our calculation of the Casimir energies of a string with piecewise constant density and of two perfectly conduting concentric cylinders, of similar radius, reproduce the results previously published.
\end{abstract}
\maketitle

\section{Introduction}
\label{sec:intro}

This paper deals with the calculation of the spectral zeta function
\beq
Z(s) = \sum_{n=1}^\infty \frac{1}{E_n^s}  \ ,
\label{spectral}
\eeq
associated with the eigenvalues $E_n$ of the inhomogeneous Helmholtz equation on a $d$-dimensional domain $\Omega$
\beq
(-\Delta_d) \psi_n(x_1, \dots, x_d) = E_n \Sigma(x_1, \dots,x_d) \psi_n(x_1, \dots ,x_d) \ , 
\label{Helmholtz_1}
\eeq
where $\Delta_d \equiv \frac{\partial^2}{\partial_{x_1}^2} + \dots + \frac{\partial^2}{\partial_{x_d}^2}$ is the 
laplacian operator in $d$-dimensions and $\psi_n(x_1, \dots , x_d)$ are its eigenfunctions (the index $n$ refers 
to the entire set of quantum numbers that specify a solution.). $\Sigma(x_1, \dots,x_d)$ is a density 
($\Sigma(x_1, \dots,x_d)>0$ for $\vec{x} \in \Omega$). 

In general, spectral zeta functions can be built using the eigenvalues associated to a given Schr\"odinger operator: for example Voros has studied the zeta function of anharmonic oscillators~\cite{Voros80,Voros99}, Steiner has obtained sum rules for confinement potentials of the form $V(r) = g r^p$, with $p>0$ and $g>0$~\cite{Steiner85}, Berry has studied 
the zeta function of Aharonov-Bohm quantum billiards at integer values of $s$~\cite{Berry86}, extending the results of Itzykson, Moussa and Luck\cite{Itzykson86}, who had derived an explicit expression for the sum of inverse powers of the Laplacian on  a simply connected region with Dirichlet boundary conditions at the border;  this problem has been later studied by Steiner \cite{Steiner87}, obtaining exact expressions for the spectral zeta function at positive integer values, and by Elizalde, Leseduarte and Romeo\cite{Elizalde93}, who derived exact sum rules for the zeros of the 
Bessel functions $J_\nu$ using a recursive formula (these sum rules were already obtained by in \cite{Steiner85}, considering an infinite square well of radius $R$); Steiner has also obtained exact expressions for the Selberg's zeta function for compact Riemann surfaces\cite{Steiner87b} (in particular the case of the two-torus is discussed in detail in the review article \cite{Arendt09} by Arendt et al.); Crandall\cite{Crandall96} has calculated the exact values of the spectral zeta functions at integer values for several one dimensional potentials~\footnote{In particular he has showed that $Z(1)$ for the power-law oscillator obtained by Voros holds for arbitrary power-laws with positive real exponents. See also ref.~\cite{Borwein00}.}; Kvitsinsky \cite{Kvitsinsky96} has obtained sum rules for two-dimensional domains which are close to the unit disc;  Mezincescu \cite{Mezincescu00}, and later Bender and Wang\cite{Bender00}, have  used the spectral zeta functions to obtain strong evidence of the reality of the spectrum of certain $\mathcal{PT}$-symmetric hamiltonians. 

Spectral zeta functions are also useful tools in the calculation of the Casimir effect, since the Casimir energy is obtained as $1/2 \ Z(-1/2)$. The series given in eq.(\ref{spectral}) diverges in this case and one needs to analytically continue it to negative values of $s$ to obtain a finite physical value for the Casimir energy: this is the essence of zeta function regularization 
method~\cite{Hawking, Dowker, Elizalde94, Kirsten01, Elizalde12}. Applications of these techniques may be found in \cite{Elizalde89,Elizalde96, Elizalde98,Elizalde06}.

Unfortunately, the direct use of eq.(\ref{spectral}) is possible only if the eigenvalues $E_n$ are known exactly;  in this paper we develop a perturbative approach to the calculation of $Z(s)$ that  allows one to obtain explicitly the corrections to the spectral zeta function to a given order in the perturbation, bypassing eq.~(\ref{spectral}). In our calculation we assume that $\Omega$ is a $d$-cube with a density $\Sigma$ which is a perturbation of a uniform density (in two dimensions
$\Omega$ may also be obtained from the small deformation of a square, a circle or a domain where a basis is available). 

We have derived a general expression for the spectral zeta function for problems with small inhomogeneities and/or deformations, which we have worked out explicitly to second order in the perturbation. We have applied our formula to different problems, obtaining in many cases the analytic continuation to all values of $s$ (this has been generally accomplished only to first  order, and to second order in a specific example a one dimensional string). For the case of a string with piecewise constant density, we have reproduced to first order in the perturbation the results obtained by Hadasz, Lambiase and Nesterenko\cite{Lambiase00}. Although most calculations of Casimir energies for inhomogeneous systems concern the special case of piecewise constant densities\cite{Brevik90, Li91, Brevik94, Brevik96}, our formalism applies to more general densities. In particular, we have observed that the Casimir energy of an inhomogeneous string, calculated to first order in perturbation theory, in some cases contains irremovable divergences, when either Dirichlet or Neumann boundary conditions 
are used, but that these singularities cancel in the sum of the two zeta functions corresponding to the different boundary conditions.

This paper is organized as follows: in section \ref{sec:rspt} we briefly introduce the Rayleigh-Schr\"odinger perturbation theory for a general hermitian operator; in section \ref{sec:zeta}  we obtain an explicit formula for the spectral zeta function using the perturbative expansion of section \ref{sec:rspt};  in section \ref{sec:applications} we discuss several applications in one, two  and three dimensions;  finally, in section \ref{conclusions} we draw our conclusions.

\section{Rayleigh-Schr\"odinger perturbation theory}
\label{sec:rspt}

Rayleigh-Schr\"odinger perturbation theory is a standard tool of quantum mechanics: it allows to obtain perturbative expressions for the eigenvalues and the eigenstates of a hermitian operator 
$\hat{C} \equiv \hat{A} + \eta \hat{B}$, defined as the sum of a solvable hermitian operator $\hat{A}$ and a of perturbation $\eta \hat{B}$ ($|\eta|\ll 1$), which typically does not commute with $\hat{A}$. 

In this framework, using the basis of the eigenstates of $\hat{A}$, one obtains an explicit expression for the eigenvalues of $\hat{C}$, as a power series in $\eta$, which to second order reads
\beq
c_n &\approx& a_n + \eta \langle n | \hat{B} | n \rangle + \eta^2 \sum_{k \neq n} \frac{|\langle n | \hat{B} | k \rangle|^2}{a_n-a_k} 
+ \dots \ . 
\label{pert}
\eeq
The formula above assumes that $|n\rangle$ is a non-degenerate eigenstate of $\hat{A}$.

Assuming  that for an arbitrary small (but finite) $\eta$, $\eta \hat{B}$ is a perturbation {\sl for all the states} belonging to the  spectrum of $\hat{C}$, we may write the trace of $\hat{C}^s$ as a power series in $\eta$ using the explicit perturbative expression for $c_n$:
\beq
{\rm Tr} \ \hat{C}^s = \sum_n  \left[ a_n^s + \eta s a_n^{s-1} \langle n | \hat{B} | n \rangle + \eta^2 \frac{s (s-1)}{2} 
a_n^{s-2} \langle n | \hat{B} | n \rangle^2 + 
s \frac{\eta^2 }{2} \sum_n \sum_{k \neq n}   \frac{a_n^{s-1} -a_k^{s-1}}{a_n-a_k} |\langle n | \hat{B} | k \rangle|^2 + 
O\left[\eta^3\right]  \right] \ ,
\label{tracec}
\eeq
where the last term has been symmetrized with respect to the summation indices. Notice that this term is well defined even when $|k\rangle$ and $|n \rangle$ are degenerate states, even though the corresponding expression in eq.(\ref{pert}) diverges in this case.

In a similar fashion one may extend these results to the operator $f(\hat{C}) \equiv f(\hat{A} + \eta \hat{B})$, obtaining
\beq
{\rm Tr} \ f(\hat{C}) &=& \sum_{n}  \left[ f(a_n) + \eta f'(a_n) \langle n | \hat{B} | n \rangle + 
\frac{\eta^2}{2} f''(a_n)  \langle n | \hat{B} | n \rangle^2 +
\frac{\eta^2}{2} \sum_{n} \sum_{k \neq n}   \frac{f'(a_n) -f'(a_k)}{a_n-a_k} 
\langle n | \hat{B} | k \rangle^2 + \dots  \right] \ .
\label{tracef}
\eeq

\section{Spectral zeta functions}
\label{sec:zeta}

We work with the spectral zeta function of Eq.(\ref{spectral}) associated with the eigenvalues $E_n$ of the hermitian operator 
$\hat{O} = \frac{1}{\sqrt{\Sigma}} (-\Delta_d ) \frac{1}{\sqrt{\Sigma}}$, on the d-dimensional cube $\Omega$  
($\Delta_d$ is the d-dimensional Laplacian and $\Sigma(x_1, \dots, x_d)$ is a positive function on $\Omega$).
The spectrum of $\hat{O}$ is bounded from below, with strictly positive eigenvalues for Dirichlet boundary conditions for $(x_1,x_2,\dots,x_d) \in \partial \Omega$.  

If $\Sigma$ is a physical density, then the eigensolutions of $\hat{O}$ provide the normal modes of an inhomogeneous d-dimensional cube; in two dimensions, $\Sigma$ may also be  obtained by conformally mapping the eigenvalue problem defined on an arbitrary two dimensional domain onto the square.

The specific form of $\hat{O}$ given above is obtained from the Helmholtz equation (\ref{Helmholtz_1}) after recasting it into the equivalent form
\beq
\frac{1}{\sqrt{\Sigma}} (-\Delta_d) \frac{1}{\sqrt{\Sigma}} \phi_n(x_1, \dots, x_d) = E_n \phi_n(x_1, \dots, x_d) \ ,
\label{Helmholtz_2}
\eeq
where $\phi_n(x_1, \dots, x_d) \equiv \sqrt{\Sigma} \ \psi_n(x_1, \dots, x_d)$. 
The advantage of working with eq.(\ref{Helmholtz_2}) rather than with eq.~(\ref{Helmholtz_1}) resides in the 
manifestly hermitian form of the operator $\hat{O} \equiv \frac{1}{\sqrt{\Sigma}} (-\Delta_d) \frac{1}{\sqrt{\Sigma}}$.

Alternatively, one may obtain a still equivalent equation
\beq
\frac{1}{\sqrt{(-\Delta_d)}}  \Sigma(x_1, \dots, x_d) \frac{1}{\sqrt{(-\Delta_d)}} \xi_n(x_1, \dots, x_d) = \frac{1}{E_n} \xi_n(x_1, \dots, x_d) \ , 
\label{Helmholtz_3}
\eeq
where $\xi_n(x_1, \dots, x_d) \equiv \sqrt{(-\Delta_d)} \ \psi_n(x_1, \dots, x_d)$. 

In this case one works with the hermitian operator 
$\hat{Q} \equiv \frac{1}{\sqrt{(-\Delta_d)}}  \Sigma(x_1, \dots, x_d) \frac{1}{\sqrt{(-\Delta_d)}}$~\footnote{Another
possibility would be working with the operator $\hat{O}^{-1} = \sqrt{\Sigma} (-\Delta_d)^{-1} \sqrt{\Sigma}$, which being the 
inverse of $\hat{O}$ is isospectral to $\hat{Q}$.}.
Notice that the eigenvalues of eq.~(\ref{Helmholtz_3}) are the reciprocal of the eigenvalues of eqs.(\ref{Helmholtz_1}) and 
(\ref{Helmholtz_2}).

In $d$ dimensions, the asymptotic behavior of the energies of eq.~(\ref{Helmholtz_1}) for $n \gg 1$ is described by Weyl's law, 
$E_n  \approx 4 \pi \left( \Gamma(1+d/2) \ \frac{n}{V} \right)^{2/d}$, where $n$ is the number of states with energy smaller 
than $E_n$. As a result,  the series $\sum_{n=1}^\infty \frac{1}{E_n^s}$ will converge to a finite value only for $s>d/2$.

Using the invariance of the trace with respect to unitary transformations, for $s>d/2$, we may express the spectral zeta function directly in terms of the diagonal matrix elements of the operators  $\hat{O}^{-s}$ and $\hat{Q}^{s}$:
\beq
Z (s) &=& \sum_{n_1,\dots,n_d} \langle n_1, \dots, n_d | \hat{O}^{-s} |n_1, \dots, n_d \rangle 
=  \sum_{n_1,\dots,n_d} \langle n_1, \dots, n_d | \hat{Q}^s |n_1, \dots, n_d \rangle \ , \ \ \ s>d/2
\label{zeta2}
\eeq
where $|n_1, \dots, n_d \rangle$ is an eigenstate of $(-\Delta_d)$ on the d-cube $\Omega$ and $\epsilon_{n_1,\dots, n_d}$ the corresponding eigenvalue (when possible we will use the compact notation $|n\rangle$ and $\epsilon_{n}$ to refer to $|n_1, \dots, n_d \rangle$ and to $\epsilon_{n_1,\dots, n_d}$ ). For integer values of $s$, with $s>d/2$, eq.(\ref{zeta2}) provides exact sum rules.

We will now discuss the case in which  $\Sigma(x_1, \dots, x_d) = 1 + \delta \Sigma(x_1, \dots, x_d)$,
with $|\delta \Sigma | \ll 1$, corresponding to a small perturbation of a uniform problem; taking into 
account this fact we write:
\beq
\hat{Q}^s = \left( \frac{1}{(-\Delta_d)} + \frac{1}{\sqrt{-\Delta_d}} \delta \Sigma   \frac{1}{\sqrt{-\Delta_d}} \right)^s \nonumber 
\eeq
and treat the second term in the parenthesis as a perturbation.

After identifying $\hat{A} = (-\Delta_d)^{-1}$ and $\hat{B} = (-\Delta_d)^{-1/2} \delta\Sigma  (-\Delta_d)^{-1/2}$ we 
then obtain
\beq
Z(s) &=& {\rm Tr}  \hat{Q}^s =  \sum_n \left( \frac{\langle n | \Sigma | n \rangle}{\epsilon_n}  \right)^s 
- \frac{s}{2}  \sum_n \sum_{k \neq n} \frac{\epsilon_k^{1-s}-\epsilon_n^{1-s}}{\epsilon_k-\epsilon_n} 
\langle n | \delta \Sigma | k \rangle^2 +  O\left[\delta\Sigma^3\right] 
\label{spectralzeta}
\eeq
where $ O\left[\delta\Sigma^3\right]$ refers only to terms containing off-diagonal matrix elements, 
since the purely diagonal contributions have been included to all orders. 
We define the first term in the expression above 
\beq
Z^{(diag)}(s) &\equiv&   \sum_n \left( \frac{\langle n | \Sigma | n \rangle}{\epsilon_n}  \right)^s  \nonumber \ ,
\eeq
to be the component of $Z(s)$ which only contains diagonal matrix elements.

There is an important aspect concerning eq.(\ref{spectralzeta}): since the matrix elements of $\langle n | \hat{B} | k\rangle$ 
are always finite in the limit $k,n \rightarrow \infty$~\footnote{It is easy to check that 
$\lim_{n \rightarrow \infty} \langle n | \delta \Sigma | n \rangle = \frac{1}{(2L)^d} \int_{\Omega_d} \delta\Sigma(x) d^dx$ and $\lim_{n \rightarrow \infty} \langle n | \delta \Sigma | k \rangle = 0$ for a fixed $k$, with $k \neq n$; in ref.~\cite{Amore11}  explicit expressions for these matrix elements for Dirichlet bc in one dimension are derived.}, the perturbative hierarchy in the expression for $Z(s)$ is not destroyed by an unbounded growth of the perturbative corrections with respect to the quantum numbers and therefore eq.(\ref{spectralzeta}) provides a genuine perturbative expansion in this case.  

On the other hand, we can easily find an example where the perturbative expansion for the spectral zeta function breaks down: in the case of the dimensional anharmonic oscillator, with hamiltonian $\hat{H} = \frac{\hat{p}^2+\hat{x}^2}{2} + \lambda x^4$ ($0 < \lambda \ll 1$), the perturbative corrections to the eigenvalues of $\hat{H}$ are polynomials in the quantum number $n$ of increasing order 
\beq
E_n = (n+1/2)  + \lambda \left(\frac{3}{2} n^2 + \frac{3}{2} n + \frac{3}{4} \right) + \lambda^2
\left(-\frac{17}{4}n^3-\frac{51}{8}n^2-\frac{59}{8} n -\frac{21}{8}\right) + \dots
\eeq
and therefore for any arbitrary small $\lambda$ the expansion breaks down for sufficiently large $n$. 
Therefore, for any finite value of $\lambda$ one cannot apply perturbation theory to all the spectrum of $\hat{H}$. Notice however that our formalism applies to the hamiltonian  $\hat{H} = \frac{\hat{p}^2}{2} + \eta \ V(x)$ ($0 < \eta \ll 1$), where  
\beq
V(x) = \left\{ \begin{array}{cc}
v(x) & x_{min} \leq x \leq x_{max} \\
\infty & x<x_{min}  \ , \ x> x_{max} \\
\end{array}\right. \ ,
\nonumber
\eeq
and $v(x)$ is a potential bounded from below. In this case the asymptotic behavior of the spectrum is determined by the confinement of the particle between the infinite walls and does not depend on the perturbation. This property is sufficient to make $\eta V(x)$ a perturbation for all the spectrum, for a given infinitesimal $\eta$.

Let us now consider the operator:
\beq
\hat{W} \equiv \hat{Q}^{-1} = \sqrt{(-\Delta_d)}  \frac{1}{\Sigma(x_1, \dots, x_d)} \sqrt{(-\Delta_d)} \ ,
\eeq
which is isospectral to $\hat{O} = \frac{1}{\sqrt{\Sigma}} (-\Delta_d) \frac{1}{\sqrt{\Sigma}}$ used in ref.\cite{Amore10}. 
The heat kernel of $\hat{W}$ is defined as
\beq
K(t) = \sum_{n} e^{-\lambda_n t} \ ,
\eeq
where $\lambda_n$ are the eigenvalues of $\hat{W}$.

In order to apply eq.(\ref{tracef}) we write
\beq
\hat{W} = (-\Delta_d) +  \sqrt{(-\Delta_d)}  \left[\frac{1}{\Sigma(x_1, \dots, x_d)} -1 \right]\sqrt{(-\Delta_d)}
\eeq
and identify $\hat{A} = (-\Delta_d)$ and $\hat{B} = \sqrt{(-\Delta_d)}  \left[\frac{1}{\Sigma(x_1, \dots, x_d)} -1 \right]\sqrt{(-\Delta_d)}$.

Using eq.(\ref{tracef}) the heat kernel reads
\beq
K(t)   \approx \sum_n e^{- (\epsilon_n + \langle n | \hat{B} | n \rangle) t} -\frac{t}{2} \sum_n \sum_{k \neq n} 
\left( \frac{e^{-t \epsilon_n}-e^{-t \epsilon_k}}{\epsilon_n -\epsilon_k}\right)  \langle n | \hat{B} | k\rangle^2 + \dots \ .
\label{HeatKernel}
\eeq

Notice that for $t\rightarrow 0^+$ we have
\beq
-\frac{t}{2} \sum_n \sum_{k \neq n} 
\left( \frac{e^{-t \epsilon_n}-e^{-t \epsilon_k}}{\epsilon_n -\epsilon_k}\right)  \langle n | \hat{B} | k\rangle^2  \rightarrow 
\frac{t^2}{2} \sum_n \left[\langle n | \hat{B}^2 | n \rangle -\langle n | \hat{B} | n \rangle^2 \right] \ ,
\eeq
using the completeness of the basis; in this limit we have the perturbative expansion for the heat kernel:
\beq
K(t) \approx \sum_n e^{-\epsilon_n t} \left[ 1 - t \langle n | \hat{B} | n \rangle + \frac{t^2}{2} \langle n | \hat{B}^2 | n \rangle +\dots  \right] \ .
\eeq

The spectral zeta functions associated with the eigenvalues of $\hat{W}$ may be obtained from the heat kernel as
\beq
Z(s) = \frac{1}{\Gamma(s)} \int_0^\infty t^{s-1} \ K(t) \  dt \ .
\label{HeatZeta}
\eeq

\section{Applications}
\label{sec:applications}

In this section we consider several applications of the formulas obtained in the previous sections.

\subsection{One dimension: Inhomogeneous strings}
\label{strings}

Before discussing specific examples of inhomogeneous strings, it is useful to make some general considerations: the Casimir energy of an inhomogeneous string is obtained calculating the value of its spectral zeta function at $s=-1/2$. Assuming that the density of the string is a perturbation of a constant density ($\Sigma(x) \approx 1 + \delta\Sigma$) and working to first order in perturbation theory, we have $Z(s) \approx \sum_{n=1}^\infty  \frac{1 + s \langle n | \delta \Sigma | \rangle}{\epsilon_n^s}$, where $\epsilon_n$ are the eigenvalues of the homogeneous string subject to specific boundary conditions and $| n \rangle$ are its eigenstates.

The analytic continuation of this series allows one to obtain the Casimir energy of the string, directly evaluating $Z(-1/2)$: however, if $Z(s)$ contains terms which are proportional to $\zeta(2s+2)$, the corresponding $Z(-1/2)$, calculated to first order, diverges, making the physical interpretation problematic. Previous examples of Casimir energies containing irremovable divergences are known: for example, Sen calculated the Casimir energy of a circle in two dimensions, finding that its Casimir energy is infinite~\cite{Sen81a, Sen81b}; Bender and Milton have calculated  the Casimir energy of a massless scalar field in a hyperspherical shell in D spatial dimensions
and observed that it diverges  when D was a positive even integer~\cite{Bender94}.

It is easy to see that for an inhomogeneous string, to first order in the perturbative expansion a divergent Casimir energy is obtained when the  matrix element $\langle n |\delta\Sigma |n\rangle$ contains a term going as $1/n^2$. In the case of Dirichlet boundary conditions, this matrix element may be expressed in terms of the series~\cite{Amore11}
\beq
\langle n | \delta\Sigma | n \rangle &=& \frac{1}{2L} \int_{-L}^{+L} \delta\Sigma(x) dx - \sum_{k=0}^\infty \frac{(2L)^{2k+1}}{(2 \pi n)^{2k+2}} (-1)^k \left[ \delta\Sigma^{(2k+1)}(L)  - \delta\Sigma^{(2k+1)}(-L) \right] , 
\label{eq_dir_1}
\eeq
where $\delta\Sigma^{(2k+1)}(x) \equiv \frac{d^{2k+1}\delta\Sigma}{dx^{2k+1}}$; therefore the 
Casimir energy of an inhomogeneous string is finite only if $\delta\Sigma'(x)|_{x=L}= \delta\Sigma'(x)|_{x=-L}$. This condition is fulfilled if the perturbation is either an odd function of $x$ or if $\delta\Sigma(x)$ is flat at the border: $\delta\Sigma'(x)|_{x=L}=\delta\Sigma'(x)|_{x=-L}=0$.

Therefore the Casimir energy of the inhomogeneous string contains an irremovable divergence when none of these conditions is met; on the other hand, irremovable divergences may still show up to higher orders, even when these are absent to first order, as we will see in an example.

\subsubsection{Casimir energy of a string with piecewise constant density}
The Casimir energy of a string with piecewise constant density and four different sets of boundary conditions
at its ends (Dirichlet-Dirichlet, Neumann-Neumann, Dirichlet-Neumann and Neumann-Dirichlet) has been calculated in ref.\cite{Lambiase00};
the density of the string is 
\beq
\Sigma(x) &=& \left\{ \begin{array}{ccc}
\rho_1 & , & -L < x \leq x_0 \\
\rho_2 & , & x_0 < x < L \\
\end{array} \right. \nonumber \ .
\eeq

To uniform our notation to that of ref.\cite{Lambiase00}, we define $\rho_{1,2} \equiv 1/\upsilon_{1,2}^2$, $R \equiv 2L$,
$r \equiv x_0 + R/2$ and $E \equiv \omega^2$. 

$\alpha$, $\beta$ and $\delta \upsilon$ are defined as in ref.\cite{Lambiase00}: 
\beq
\delta \upsilon \equiv \frac{\upsilon_1-\upsilon_2}{\upsilon_1+\upsilon_2} \ \ , \ \ \alpha \equiv \frac{r}{\upsilon_1} + \frac{R-r}{\upsilon_2} \ \ , \ \
\beta \equiv \frac{r}{\upsilon_1} - \frac{R-r}{\upsilon_2} \nonumber \ .
\eeq

For Dirichlet boundary conditions, the exact eigenvalues are the solutions of the trascendental equation 
\beq
\sin \alpha \omega + \delta\upsilon \sin \beta\omega = 0 \ , \nonumber
\eeq
which corresponds to the first of eqns. (2.4) of ref.~\cite{Lambiase00}.

For a string of arbitrary density, the energies go asymptotically as $E_n \approx \frac{n^2\pi^2}{\sigma(L)^2}$, where
$\sigma(L) \equiv \int_{-L}^{+L} \sqrt{\rho(x)} dx$ (see for instance \cite{BO78}): in the present
case $\sigma(L) = \alpha$ and therefore $E_n \approx n^2\pi^2/\alpha^2$ for $n \rightarrow \infty$, which is the behavior of the spectrum of a uniform string with constant density $\bar{\Sigma} = \frac{\sigma(L)^2}{R^2} = \frac{\alpha^2}{R^2}$.

We may therefore write
\beq
\Sigma(x) \approx \bar{\Sigma}  +\delta\Sigma(x) \nonumber
\eeq
and apply our formulas to first order in $\delta\Sigma$, for which only diagonal matrix elements of the perturbation are needed. 

For Dirichlet boundary conditions, the basis is
\beq
\psi_n^{(D)}(x) = \sqrt{\frac{1}{L}} \ \sin \frac{n \pi (x+L)}{2L} \ ,  \nonumber
\eeq
with $n=1,2, \dots$. 

Correspondingly, the eigenvalues of the negative laplacian with Dirichlet bc are
\beq
\epsilon_n^{({\rm DD})} = \frac{n^2\pi^2}{4L^2} \nonumber \ .
\eeq

The matrix elements of $\Sigma$ in this basis are
\begin{eqnarray}
\left. \langle n | \Sigma | n' \rangle\right|^{({\rm DD})} = \left\{ \begin{array}{ccc}
\frac{\pi  n \left(\upsilon_1^2 (L-r)+\upsilon_2^2 (L+r)\right)+L (\upsilon_1-\upsilon_2) (\upsilon_1+\upsilon_2) \sin
   \left(\frac{\pi  n (L+r)}{L}\right)}{2 \pi  L n \upsilon_1^2 \upsilon_2^2} &  , &  n=n'\\
-\frac{(\upsilon_1^2-\upsilon_2^2)\left((n+{n'}) \sin \left(\frac{\pi  (L+r) (n-{n'})}{2
   L}\right)+({n'}-n) \sin \left(\frac{\pi  (L+r) (n+{n'})}{2 L}\right)\right)}{\pi  \upsilon_1^2 \upsilon_2^2
   (n^2-{n'}^2) } &  , &  n \neq n' \\   
\end{array}
\right. \ . \nonumber 
\end{eqnarray}
Notice that $\left. \langle n | \Sigma | n \rangle\right|^{({\rm DD})}$ does not contain terms going like $1/n^2$ for $n \rightarrow \infty$ and therefore 
we expect that the Casimir energy of this string will be finite.

A simple calculation, carried out to order $\delta\upsilon$, yields
\beq
Z^{({\rm DD})}(s) &\approx& \bar{\Sigma}^s \sum_{n=1}^\infty \left\{ \frac{1}{{\epsilon^{({\rm DD})}_n}^s} + s  
\frac{\left.\langle n | \left[ \Sigma /\bar{\Sigma} - 1 \right] | n \rangle\right|^{({\rm DD})} }{{\epsilon^{({\rm DD})}_n}^s} \right\} \nonumber \\
&=& \pi ^{-2 s-1} \alpha ^{2 s} \left[\pi  \zeta (2 s)  + i \ s \ \delta\upsilon    \left({\rm Li}_{2 s+1}\left(e^{-\frac{i (\alpha +\beta ) \pi
   }{\alpha }}\right)-{\rm Li}_{2 s+1}\left(e^{\frac{i (\alpha +\beta
   ) \pi }{\alpha }}\right)\right)\right] \ ,  \nonumber 
\eeq
where $\epsilon_n \equiv n^2\pi^2/4L^2$ and ${\rm Li}_{\nu}(z) \equiv \sum_{k=1}^\infty \frac{z^k}{k^\nu}$ is the polylogarithmic function.

We observe that this equation provides an analytic continuation of the series to negative values of $s$; in particular, for $s= -1/2$ this formula returns the Casimir energy of the string, which reads 
\beq
E_C^{({\rm DD})} = \frac{1}{2} Z^{({\rm DD})}(-1/2) = -\frac{\pi }{24 \alpha }+\frac{\delta\upsilon \tan \left(\frac{\pi  \beta }{2 \alpha }\right)}{4 \alpha } \ .
\label{piecewise}
\eeq

The same result can be obtained regularizing the series, evaluated at $s=-1/2$, with a cutoff function, $f(a,n) \equiv e^{-a n}$, and taking the limit $a \rightarrow 0^+$ at the  end of the calculation:
\beq
E_C^{({\rm DD})} &=& \frac{1}{2} Z^{({\rm DD})}(-1/2) = \frac{1}{2 \sqrt{\bar{\Sigma}}} \sum_{n=1}^\infty \sqrt{\epsilon_n} 
\left[ 1 - \frac{1}{2} \langle n | \left. \left[ \Sigma /\bar{\Sigma} - 1 \right] | n \rangle \right|^{({\rm DD})} \right] e^{-a n} \nonumber \\
&=& \frac{\pi }{2 a^2 \alpha }+\frac{\delta\upsilon \tan \left(\frac{\pi  \beta }{2 \alpha }\right)}{4\alpha }-\frac{\pi }{24 \alpha } \ \ \  , \ \ \ a \rightarrow 0^+ \ \ . \nonumber 
\eeq
Notice that the finite part of this expression agrees with the result of eq.(\ref{piecewise}).

We can compare this result with the analogous result of eq.(3.5) and (3.6) of ref.\cite{Lambiase00}; taking $\delta\upsilon \ll 1$ and
calculating (3.5) to order $\delta\upsilon$ we obtain the result of our eq.(\ref{piecewise}).

We may also calculate the exact value of the spectral zeta function for this string at $s=1$:
\beq
Z^{({\rm DD})}(1)  = \sum_{n=1}^\infty \frac{1}{E_n^{({\rm DD})}} =  
\frac{(\alpha^2 -\beta^2 \delta\upsilon^2)}{6 (1- \delta\upsilon^2)} + i \delta\upsilon
\frac{(\alpha +\beta  \delta\upsilon)^2}{\pi ^3\left(\delta\upsilon^2-1\right)^2}  \left(\text{Li}_3\left(e^{-\frac{i (\alpha +\beta )
   (\delta\upsilon+1) \pi }{\alpha +\beta  \delta\upsilon}}\right)-\text{Li}_3\left(e^{\frac{i (\alpha
   +\beta ) (\delta\upsilon+1) \pi }{\alpha +\beta  \delta\upsilon}}\right)\right) \nonumber \ .
\eeq
This expression holds to all orders in $\delta\upsilon$.

The calculation for the cases of Neumann-Neumann, Neumann-Dirichlet, Dirichlet-Neumann and periodic-periodic boundary conditions can be done in a similar fashion. Here we just report the results.

The spectral zeta functions for these cases, calculated to first order, are
\beq
Z^{({\rm NN})}(s)  &=& \pi ^{-2 s} \alpha ^{2 s} \zeta (2 s) + 
i \delta\upsilon (2 \pi )^{-2 s-1} s e^{-\frac{i \pi  \beta }{\alpha }} \alpha ^{2 s} 
\left(\Phi\left(e^{-\frac{2 i \beta  \pi }{\alpha }},2 s+1,\frac{1}{2}\right) \right. \nonumber\\
&-& \left. e^{\frac{2 i \pi  \beta}{\alpha }} 
\Phi \left(e^{\frac{2 i \beta  \pi }{\alpha }},2 s+1,\frac{1}{2}\right) 
-e^{\frac{i \pi  \beta }{\alpha }} \text{Li}_{2 s+1}
\left(e^{-\frac{2 i \beta  \pi }{\alpha }}\right)+e^{\frac{i \pi  \beta }{\alpha }} \text{Li}_{2s+1}\left(e^{\frac{2 i \beta  \pi }{\alpha }}\right)\right) \ , \nonumber \\
Z^{({\rm DN})}(s) &=& \left(4^s-1\right) \pi ^{-2 s} \alpha ^{2 s} \zeta (2 s) \nonumber \\
&+& \delta\upsilon \pi ^{-2 s-1} s \alpha ^{2 s} \left(e^{-\frac{i \pi  \beta }{2 \alpha }} \Phi \left(-e^{-\frac{i \beta  \pi}{\alpha }},2 s+1,\frac{1}{2}\right)
+e^{\frac{i \pi  \beta }{2 \alpha }} \Phi \left(-e^{\frac{i \beta  \pi }{\alpha}},2 s+1,
\frac{1}{2}\right)\right) \ , \nonumber \\
Z^{({\rm PP})}(s) &=& 2^{1-2 s} \pi ^{-2 s} \alpha ^{2 s} \zeta (2 s) \ , \nonumber
\eeq
where $\Phi\left(z,s,a\right) \equiv \sum_{k=0}^\infty \frac{z^k}{(a+k)^s}$ is the Lerch transcendent  function. The case of Neumann-Dirichlet bc is simply obtained from the case Dirichlet-Neumann bc with the substitution $\delta\upsilon \rightarrow - \delta\upsilon$.

The Casimir energies for these cases are then
\beq
E_C^{({\rm NN})}  &=&  -\frac{\pi }{24\alpha } -\frac{\delta\upsilon \tan \left(\frac{\pi  \beta }{2 \alpha }\right)}{4 \alpha } \ ,  \nonumber \\
E_C^{({\rm DN})}  &=& \frac{\pi }{48 \alpha }-\frac{\delta\upsilon \sec \left(\frac{\pi  \beta }{2 \alpha }\right)}{4 \alpha } \ , \nonumber \\
E_C^{(ND)}  &=& \frac{\pi }{48 \alpha } +\frac{\delta\upsilon \sec \left(\frac{\pi  \beta }{2 \alpha }\right)}{4 \alpha }  \ , \nonumber \\
E_C^{({\rm PP})}  &=& - \frac{\pi}{6\alpha} \ . \nonumber 
\eeq

These results agree to order $\delta\upsilon$ with the expressions calculated in ref.\cite{Lambiase00}, with the exception of the PP case which 
was not studied there.

Exact sum rules can also be obtained using eq.(\ref{zeta2}):
\beq
Z^{({\rm NN})}(1)  &=&  \frac{(\alpha^2 -\beta^2 \delta\upsilon^2)}{6 (1- \delta\upsilon^2)}-i \delta\upsilon 
\frac{(\alpha +\beta  \delta\upsilon)^2}{\pi ^3\left(\delta\upsilon^2-1\right)^2}  \left(\text{Li}_3\left(e^{-\frac{i (\alpha +\beta )
(\delta\upsilon+1) \pi }{\alpha +\beta  \delta\upsilon}}\right)-
\text{Li}_3\left(e^{\frac{i (\alpha+\beta ) (\delta\upsilon+1) \pi }{\alpha +\beta  \delta\upsilon}}\right)\right) \ ,   \nonumber \\
Z^{({\rm DN})}(1) &=& \frac{(\alpha -\beta  \delta\upsilon) (\alpha +\beta  \delta\upsilon)}{2-2 \delta\upsilon^2} + 
\frac{i \delta\upsilon (\alpha +\beta  \delta\upsilon)^2 }{\pi ^3
   \left(\delta\upsilon^2-1\right)^2} \nonumber \\
&\cdot& \left(e^{-\frac{i \pi  (\delta\upsilon+1) (\alpha +\beta )}{2 (\alpha +\beta 
   \delta\upsilon)}} \Phi \left(e^{-\frac{i (\alpha +\beta ) (\delta\upsilon+1) \pi }{\alpha +\beta 
   \delta\upsilon}},3,\frac{1}{2}\right)-e^{\frac{i \pi  (\delta\upsilon+1) (\alpha +\beta )}{2 (\alpha +\beta  \delta\upsilon)}} \Phi
   \left(e^{\frac{i (\alpha +\beta ) (\delta\upsilon+1) \pi }{\alpha +\beta  \delta\upsilon}},3,\frac{1}{2}\right)\right) \ ,\nonumber \\ 
Z^{({\rm PP})}(1) &=& \frac{\alpha ^2-\beta ^2 \delta\upsilon^2}{12-12 \delta\upsilon^2} \nonumber \ .
\eeq

Notice that
\beq
Z^{({\rm PP})}(1) =\frac{1}{4}  \left(Z^{({\rm DD})}(1) + Z^{({\rm NN})}(1)\right) \nonumber \ .
\eeq

\subsubsection{A slightly inhomogeneous string}

We consider a string of length $2L$ ($|x| \leq L$) and  with density
\beq
\Sigma(x) = 1 +\eta \delta\Sigma \equiv 1 + \eta \sin \frac{\pi x}{2L} \ , \nonumber
\eeq
where $|\eta| < 1$ (this condition is needed to enforce $\Sigma(x) >0$ for $x \in (-L,L)$).

In this case the matrix elements of $\delta\Sigma$ in the basis of the homogeneous string are particularly simple
\beq
\langle n | \delta \Sigma | m \rangle = - \frac{1}{2} \left[\delta_{m,n+1}+\delta_{m,n-1}\right] \ , \nonumber 
\eeq
and allow one to calculate exactly the value of the spectral zeta function of the string at the
first few positive integers values of $s$:
\beq
Z(1) &=& \frac{2}{3} L^2 \nonumber \\
Z(2) &=& \frac{8 L^4}{45}   + \frac{8 \eta ^2 L^4}{3 \pi ^2}-\frac{24 \eta ^2 L^4}{\pi ^4} \nonumber \\
Z(3) &=& \frac{64 L^6}{945} + \frac{16 \eta ^2 L^6}{15 \pi ^2}+\frac{64 \eta ^2 L^6}{\pi ^4}-\frac{720 \eta ^2 L^6}{\pi^6}
\nonumber  \ .
\eeq

Using our eq.(\ref{tracef}) we may calculate the spectral zeta function of this string to second order:
\beq
Z(s) &=&  4^s \left(\frac{L}{\pi}\right)^{2 s} \zeta(2s) + \eta^2 s 4^{s-1} \left(\frac{L}{\pi}\right)^{2 s} \sum_{n=1}^\infty 
\frac{\left(n^{2-2 s}-(n+1)^{2-2 s}\right)}{2 n+1} + O\left[\eta^3\right]  \nonumber  \ .
\eeq

Let us define:
\beq
\Psi^{(1)}(a) &\equiv& \sum_{n=1}^\infty \frac{n^a}{2 n+1} = \sum_{j=0}^\infty \frac{(-1)^j}{2^{j+1}} \ \zeta(j+1-a) 
= \frac{1}{3} + \sum_{j=0}^\infty \frac{(-1)^j}{2^{j+1}} \ \left( \zeta(j+1-a) -1 \right) \nonumber \ , \\
\Psi^{(2)}(a) &\equiv& \sum_{n=1}^\infty \frac{(n+1)^a}{2 n+1} = \sum_{j=0}^\infty \frac{1}{2^{j+1}} \ \left(\zeta(j+1-a)-1\right) \nonumber \ ,
\eeq
and
\beq
\Psi(a) &\equiv& \Psi^{(1)}(a) -\Psi^{(2)}(a) =  
\frac{1}{3} - \sum_{j=0}^\infty \frac{1}{2^{2j+1}} \ \left( \zeta(2j+2-a) -1 \right) \ .
\label{psis}
\eeq
Notice that $\Psi(a)$ is singular at odd values of $a$, $a=1,3,5,\dots$ and that $\Psi(2) = \Psi(4) = \dots = 1$.

Thus the spectral zeta function reads
\beq
Z(s) &=&  4^s \left(\frac{L}{\pi}\right)^{2 s} \zeta(2s) + \eta^2 s 4^{s-1} \left(\frac{L}{\pi}\right)^{2 s} \Psi(2-2s) \ .
\nonumber
\eeq

In particular, around $s=-1/2$ we have:
\beq
Z(s) &\approx&  \frac{\pi  \eta ^2}{256 L} \ \frac{1}{(s+1/2)}  -\frac{\pi }{24 L} 
+\frac{\pi  \eta ^2}{384 L} \left(3 \log \left(\frac{8 L}{\pi }\right)+3 \gamma -31\right) +O\left( (s+1/2)^2 \right) \nonumber \ ,
\eeq 
where $\gamma \approx 0.577216$ is the Euler--Mascheroni constant.

In this case a divergence in the Casimir energy shows up to second order in perturbation theory,
since $\delta\Sigma'(x)|_{x=L}= \delta\Sigma'(x)|_{x=-L}$. Clearly the divergences which 
emerge at a given order in the perturbative expansion cannot be canceled by divergent 
contributions stemming from higher orders, given  the different dependence on the expansion parameter. The cancellation of these divergences therefore requires considering equally divergent terms which may show up in the spectral zeta function corresponding to different boundary conditions evaluated to the same order in perturbation theory.

We will now prove that the sum $Z^{({\rm DD})}(s) +Z^{({\rm NN})}(s)$ does not contain a singularity 
at $s=-1/2$ to first order in perturbation theory~\footnote{In two and three dimensions this particular combination of boundary conditions is needed to evaluate the Casimir energy of the electromagnetic field.}. To prove this point consider
\beq
Z^{(DD+NN)}(s) &\equiv& Z^{({\rm DD})}(s) +Z^{({\rm NN})}(s) \nonumber\\
&\approx& \bar{\Sigma}^s \sum_{n=1}^\infty \left\{ \frac{1}{\left(\epsilon^{({\rm DD})}_n\right)^s} + s  
\frac{\langle n | \left[ \Sigma /\bar{\Sigma} - 1 \right] | n \rangle^{({\rm DD})}}{\left(\epsilon^{({\rm DD})}_n\right)^s} \right\} \nonumber \\
&+& \bar{\Sigma}^s \sum_{u=1}^2 \sum_{n=1}^\infty \left\{ \frac{1}{\left(\epsilon^{({\rm NN})}_{n,u}\right)^s} + s  
\frac{\langle n, u | \left[ \Sigma /\bar{\Sigma} - 1 \right] | n , u\rangle^{({\rm NN})}}{\left(\epsilon^{({\rm NN})}_{n,u}\right)^s} \right\} \nonumber  \ ,
\eeq
where
\beq
\epsilon_{n,u}^{({\rm NN})} &=& \left\{  \begin{array}{ccc}
\frac{n^2\pi^2}{L^2} & , & n=0,1,2, \dots \ \ , \ \ u =1 \\
\frac{(2n-1)^2\pi^2}{4L^2} & , & n=1,2, \dots \ \ , \ \ u =2 \\
\end{array}
\right.  \nonumber 
\eeq
are the eigenvalues of the negative 1D laplacian with Neumann-Neumann boundary conditions and $\bar{\Sigma} = \sigma(L)^2/(4L^2)$.

Notice that
\beq
\epsilon_{2n}^{({\rm DD})} = \epsilon_{n,1}^{({\rm NN})} \ \ , \ \ \epsilon_{2n-1}^{({\rm DD})} = \epsilon_{n,2}^{({\rm NN})} \nonumber  \ .
\eeq

Moreover
\beq
\psi_{2n}^{({\rm DD})}(x)^2 + \psi_{n,1}^{({\rm NN})}(x)^2 = \psi_{2n-1}^{({\rm DD})}(x)^2 + \psi_{n,2}^{({\rm NN})}(x)^2 &=& \frac{1}{L} \nonumber  \ ,
\eeq
where
\beq
\psi_{n,u}^{({\rm NN})}(x) = \left\{ \begin{array}{ccc}
\frac{1}{\sqrt{2L}} &, &  n = 0 \ \ , \ \ u=1 \\
\frac{1}{\sqrt{L}} \ \cos \frac{n\pi x}{L} &, &  n = 1,2,\dots \ \ , \ \ u=1 \\
\frac{1}{\sqrt{L}} \ \sin \frac{(2n-1)\pi x}{2L} &, &  n = 1,2,\dots \ \ , \ \ u=2 \\
\end{array}\right. \ ,   \nonumber 
\eeq
are the eigenfunctions of the negative 1D laplacian with Neumann-Neumann boundary conditions.

Using these identities we may write the spectral zeta function for the one-dimensional string to first order as
\beq
Z^{(DD+NN)}(s)  &\approx& 2 \bar{\Sigma}^s  \left(1 + \frac{s}{2L}  \int_{-L}^{L} \left[ \Sigma(x) /\bar{\Sigma} - 1 \right] dx \right) \ 
\sum_{n=1}^\infty \frac{1}{{\epsilon^{({\rm DD})}_n}^s} \ ,
\label{zetaEM}
\eeq
where $\sum_{n=1}^\infty \frac{1}{{\epsilon^{({\rm DD})}_n}^s} = \pi ^{-2 s} (2L)^{2 s} \zeta (2 s)$ is the spectral zeta function of the homogeneous string.
Therefore the only singularity of the expression in eq.~(\ref{zetaEM}) is the singularity of the Riemann zeta function at $s=1/2$ and the corresponding
Casimir energy is {\sl always} finite, as anticipated.

\begin{figure}
~\bigskip\bigskip
\begin{center}
\includegraphics[width=8cm]{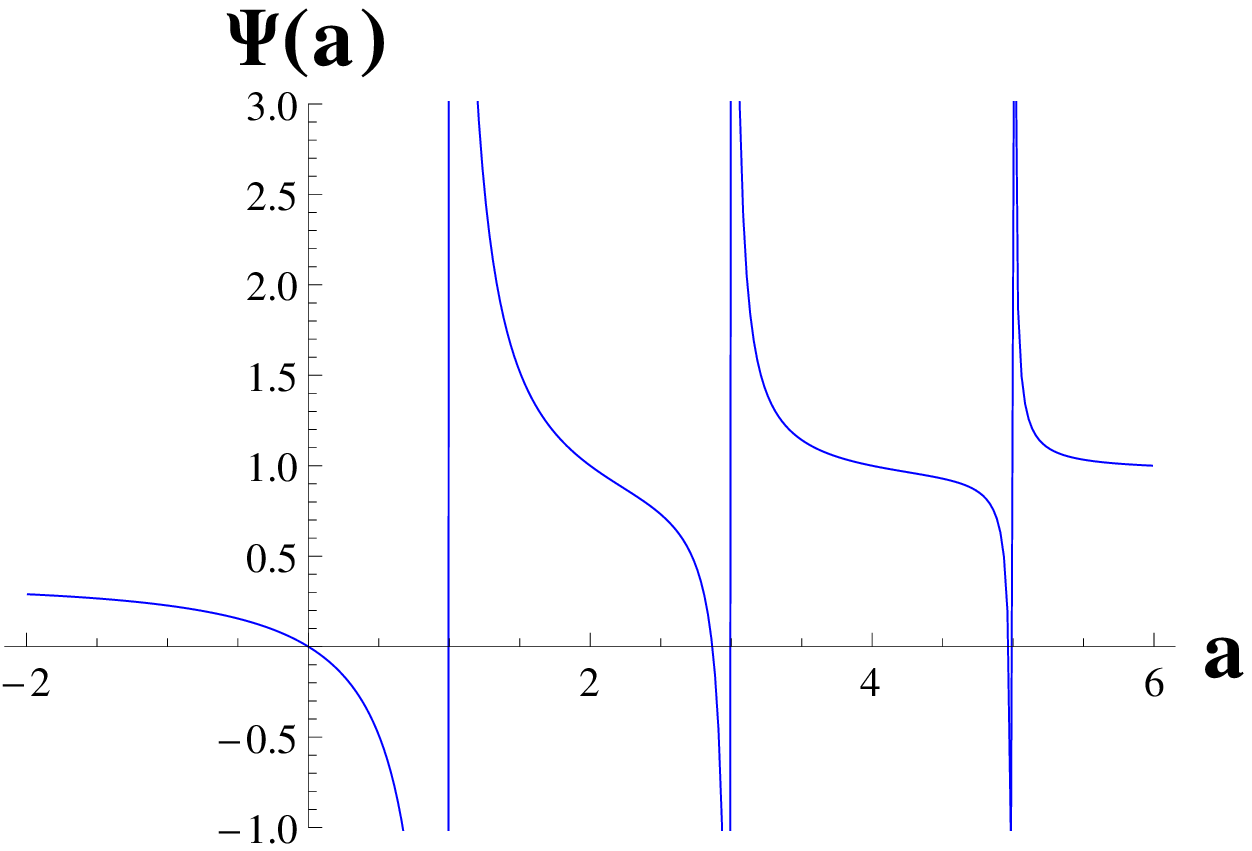}
\caption{$\Psi(a)$ of eq.(\ref{psis}).} 
\bigskip
\label{Figure}
\end{center}
\end{figure}

\subsubsection{Borg string}

We consider a string of unit length ($L=1/2$) with density
\beq
\Sigma(x) = \frac{(\alpha +1)^2}{\left(1+ \alpha \left(x+\frac{1}{2}\right)\right)^4} \ , \nonumber
\eeq
where $|x| \leq 1/2$ and Dirichlet boundary conditions at its ends. $\alpha$ is a free parameter ($\alpha >-1$).
It has been proved by Borg that this string is isospectral to a homogeneous string of the same length and unit density~\cite{Borg46}:
therefore the exact energies are
\beq
E_n = \pi^2 n^2  \nonumber
\eeq
and the corresponding spectral zeta functions are easily calculated:
\beq
Z_{Borg}(s) = \frac{\zeta(2s)}{\pi^{2s}} \ . \nonumber
\eeq

It is possible to obtain an explicit expression for the matrix elements of $\Sigma$ in the basis with Dirichlet boundary conditions;
for instance, the diagonal matrix elements read
\beq
\langle n | \Sigma | n \rangle &=& \frac{2 \pi ^2 (\alpha +1) n^2}{3 \alpha ^4}  \left[\alpha ^2  +  2 \pi  (\alpha +1) n  
   \left(\left({\rm Ci}\left(\frac{2 (\alpha +1) n \pi }{\alpha }\right)-{\rm Ci}\left(\frac{2
   n \pi }{\alpha }\right)\right) \sin \left(\frac{2 \pi  n}{\alpha
   }\right) \right.\right. \nonumber \\
   &+& \left.\left.\left({\rm Si}\left(\frac{2 n \pi }{\alpha }\right)-{\rm Si}\left(\frac{2 (\alpha
   +1) n \pi }{\alpha }\right)\right) \cos \left(\frac{2 \pi  n}{\alpha
   }\right)\right)\right] \nonumber \ ,
\eeq
where ${\rm Ci}(x)$ and ${\rm Si}(x)$ are the cosine and sine integrals.

Using these matrix elements we obtain the identity
\beq
Z_{Borg}(1) &=& \frac{1}{6} = \sum_{n=1}^\infty \frac{\langle n | \Sigma | n \rangle }{n^2\pi^2} \nonumber \\
&=& \sum_{n=1}^\infty  \frac{2 (\alpha +1)}{3 \alpha ^4}  \left[\alpha ^2+2 \pi  (\alpha +1) n \left(\left({\rm Ci}\left(\frac{2 (\alpha +1) n \pi }{\alpha }\right)-{\rm Ci}\left(\frac{2
   n \pi }{\alpha }\right)\right) \sin \left(\frac{2 \pi  n}{\alpha}\right) \right.\right. \nonumber \\
   &+& \left.\left.\left({\rm Si}\left(\frac{2 n \pi }{\alpha }\right)-{\rm Si}\left(\frac{2 (\alpha
   +1) n \pi }{\alpha }\right)\right) \cos \left(\frac{2 \pi  n}{\alpha}\right)\right)\right] \nonumber \ ,
\eeq
which we have numerically verified  with high accuracy. 

\begin{figure}
~\bigskip\bigskip
\begin{center}
\includegraphics[width=8cm]{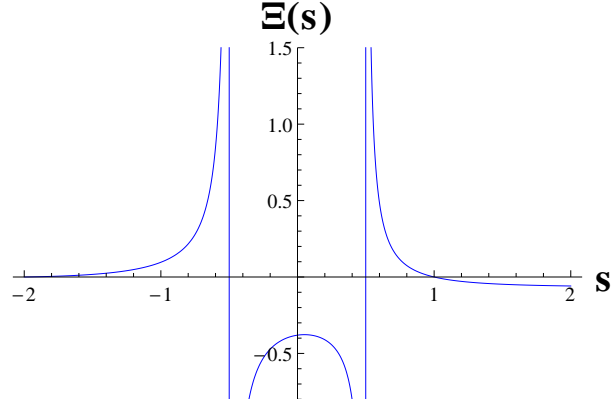}
\caption{$\Xi(s)$ of eq.(\ref{borg_analytic}) as function of $s$ (solid line).} 
\bigskip
\label{Figure2b}
\end{center}
\end{figure}

The isospectrality implies that the spectral zeta function of the Borg string is independent of $\alpha$: using this property and working to a given order in $\alpha$, we may obtain non trivial mathematical relations. For example, if we select the contribution of order $\alpha^2$ in the perturbative expression for the spectral zeta function we obtain the identity:
\begin{eqnarray}
\Xi(s) &\equiv& \sum_{k,n=1, k\neq n}^\infty \frac{k^2 n^2 \left((-1)^{k+n}-1\right) \left(n^{2-2 s}-k^{2-2s}\right)}{2\left(k^2-n^2\right)^5} 
= \frac{\pi^4}{768}  \zeta (2 s) -\frac{5 \pi^2}{256}  \zeta (2 (s+1)) \ ,
\label{borg_analytic}
\end{eqnarray}
which provides the analytic continuation of the double series to negative values  of $s$ (see Fig.\ref{Figure2b}).

We may also obtain non trivial identities working with the heat kernel: in this case we need to evaluate the matrix elements of 
$\hat{W} = \sqrt{(-\frac{d^2}{dx^2})} \frac{1}{\Sigma} \sqrt{(-\frac{d^2}{dx^2})}$:
\beq
\langle n |\hat{W} | m \rangle &=& \left\{ \begin{array}{ccc}
\frac{3 \alpha ^4}{2 \pi ^2 (\alpha +1)^2 n^2}-\frac{\left(\alpha ^2+3
   \alpha +3\right) \alpha ^2}{(\alpha +1)^2}+\frac{\pi ^2 \left(\alpha
   ^4+5 \alpha ^3+10 \alpha ^2+10 \alpha +5\right) n^2}{5 (\alpha
   +1)^2}  & , & n=m \\
\frac{256 \alpha ^2 m^4 n^4 \left(12 \alpha ^2
   \left(m^2+n^2\right)+(\alpha +1) \left(\pi ^2 (\alpha +1)^2
   \left(m^2-n^2\right)^2-12 \alpha ^2 \left(m^2+n^2\right)\right) \cos
   (\pi  (m+n))-\pi ^2 \left(m^2-n^2\right)^2\right)^2}{\pi ^4 (\alpha
   +1)^4 (m-n)^8 (m+n)^8} &  , & n \neq m
\end{array}
\right.
\eeq

If we use eq.~(\ref{HeatKernel}), to order $\alpha^2$ we obtain the identity:
\beq
\frac{256 t}{\pi^2} \sum_{n=2}^\infty \sum_{m=1}^{n-1} 
\frac{m^4 n^4 \left((-1)^{m+n}-1\right)^2 
\left(e^{-\pi ^2 n^2 t}-e^{-\pi ^2 m^2 t}\right)}{\left(m^2-n^2\right)^5} = - \frac{3}{2} t \ \vartheta _3\left(0,e^{-\pi ^2 t}\right) 
+ \frac{3 t}{2} - \frac{t}{2} \ \frac{d}{dt} \vartheta _3\left(0,e^{-\pi ^2 t}\right) \ , \nonumber \\
\label{heat}
\eeq
where $\vartheta_3(z,q) \equiv \sum_{n=-\infty}^\infty q^{n^2} e^{2n i z}$ is the Jacobi theta function. We have tested this identity 
numerically for several values of $t$.

Using eq.(\ref{HeatZeta}) inside eq.(\ref{heat}) we have
\beq
\sum_{n=1}^\infty \sum_{k\neq n=1}^\infty \frac{\left((-1)^{k+n}-1\right) (k n)^{2-2 s} \left(k^{2 s+2}-n^{2s+2}\right)}{3 \left(k^2-k^2\right)^5} 
= \frac{\pi^4}{768}  \zeta (2 s)-\frac{\pi^2}{256}  \zeta (2 (s+1)) \ .
\eeq

After combining this identity with eq.(\ref{borg_analytic}) we obtain an alternative series representation for the Riemann zeta function:
\beq
\zeta(s) = \frac{128}{\pi^4} \ \sum_{n=2}^\infty \sum_{k=1}^{n-1} 
\frac{\left((-1)^{k+n}-1\right) k^{2-s} n^{2-s} \left(-5 k^{s+2}-3 n^2 k^s+3 k^2 n^s+5 n^{s+2}\right)}{\left(k^2-n^2\right)^5} \ .
\label{zetanew}
\eeq

\subsubsection{A string with oscillating density}

We consider now  a string with rapidly oscillating density
\beq
\Sigma(x) = 2 + \eta \sin \left(2\pi (x+\bar{L)}/\varepsilon\right) \ , \nonumber
\eeq
where $|x| \leq L$ and $\varepsilon \rightarrow 0^+$. 

This particular problem has been studied by Castro and Zuazua in ref.~\cite{Castro00} and more recently by myself in ref.~\cite{Amore11}. The solutions to this problem with wavelength comparable to the typical size of the density oscillations are localized at the ends of the string~\cite{Castro00,Avellaneda92,Castro00b}. In particular, in ref.~\cite{Amore11} we have obtained both numerical and analytic approximations to the low part of the spectrum of this string (the analytic result improves the results previously obtained by Castro and Zuazua in ref.~\cite{Castro00}).

In this case the non-zero matrix elements of the density are
\beq
\langle n | \Sigma | m \rangle &=& \left\{ \begin{array}{ccc}
2-\frac{1}{2} \eta  (-1)^n \sin \left(\frac{\pi  \ell
   n}{L}\right) & , & n=m \ , \ n = 1/\bar{\varepsilon} \\
2+\frac{\bar{\varepsilon}^3 \eta  n^2 \sin \left(\frac{\pi}{\bar{\varepsilon}}\right) \sin \left(\frac{\pi 
   \ell}{\bar{\varepsilon} L}\right)}{\pi  \bar{\varepsilon}^2 n^2-\pi } 
& , &  n=m \ , \ n \neq 1/\bar{\varepsilon} \\
-\frac{8 \bar{\varepsilon}^3 \eta  m n (-1)^{m+n} \left(\cos
   \left(\frac{\pi  (L+\ell)}{\bar{\varepsilon} L}\right)-\cos (\pi 
   (m+n)) \cos \left(\frac{\pi  (L-\ell)}{\bar{\varepsilon}
   L}\right)\right)}{\pi  
   (\bar{\varepsilon}^2 (n-m)^2-4) (\bar{\varepsilon}^2 (m+n)^2-4)}
& , & n \neq m \ , \  (n\pm m) \neq \pm 2/\bar{\varepsilon}\\
\end{array}\right. \ , \nonumber
\eeq
where $\bar{\varepsilon} \equiv \varepsilon/2L$.

As in the example of the piecewise constant string we may calculate
\beq
\sigma(L) = \frac{2  L \bar{\varepsilon}}{\pi} \sqrt{\eta +2} \left[ E\left(\frac{((\bar{\varepsilon}+2) L-2 \ell) \pi }{4 \bar{\varepsilon} L}|\frac{2 \eta }{\eta+2}\right)-E\left(\frac{(\bar{\varepsilon} L-2 (L+\ell)) \pi }{4
\bar{\varepsilon} L}|\frac{2 \eta }{\eta +2}\right)\right] \ , \nonumber
\eeq
where $E \left( \phi | m \right)$ is the incomplete elliptic integral of second kind.

The asymptotic behavior of the spectrum of this string is $E_n \approx n^2 \pi^2/\sigma(L)^2$, which is also the spectrum of a homogeneous string with density $\bar{\Sigma} = \sigma(L)^2/4L^2$. 

Working to first order in $\eta$ we have
\beq
Z(s) &\approx& \bar{\Sigma}^s \sum_{n=1}^\infty \left\{ \frac{1}{\epsilon_n^s} + s  
\frac{\langle n | \left[ \Sigma /\bar{\Sigma} - 1 \right] | n \rangle}{\epsilon_n^s} \right\} \nonumber \\
&\approx& \sum_{n=1}^\infty \left\{ \left( \frac{8 L^{2}}{\pi^2}\right)^s \frac{1}{n^{2 s}} +
s \eta \frac{\bar{\varepsilon}^3   2^{3 s-1}  L^{2 s}}{\pi ^{2 s+1} }  \sin\left(\frac{\pi}{\bar{\varepsilon}}\right) 
\sin\left(\frac{\pi  \ell}{\bar{\varepsilon}L}\right)  \frac{n^{2-2 s}}{\bar{\varepsilon}^2 n^2-1} \right\}  \nonumber \ .
\eeq

This expression contains the series
\beq
\Phi(s,\bar{\varepsilon}) \equiv \sum_{n=1}^\infty \frac{n^{2-2s}}{\bar{\varepsilon}^2 n^2-1} \ ,
\label{phis}
\eeq
which we may express as
\beq
\Phi(s,\bar{\varepsilon}) &=& \sum_{n=1}^{[1/\bar{\varepsilon}]} \frac{n^{2-2s}}{\bar{\varepsilon}^2 n^2-1} +\sum_{n=[1/\bar{\varepsilon}]+1}^\infty \frac{n^{2-2s}}{\bar{\varepsilon}^2 n^2-1} 
\equiv \Phi^{(1)}(s,\bar{\varepsilon}) +\Phi^{(2)}(s,\bar{\varepsilon})
\ , \nonumber
\eeq
where $[a]$ is the integer part of $a$. 

The function $\Phi_s^{(2)}(\bar{\varepsilon})$ may be cast in the form
\beq
\Phi^{(2)}(s,\bar{\varepsilon}) &=&  \sum_{n=[1/\bar{\varepsilon}]+1}^\infty \sum_{k=0}^\infty \frac{n^{-2s-2k}}{\bar{\varepsilon}^{2k+2}}
=  \sum_{k=0}^\infty \frac{1}{\bar{\varepsilon}^{2k+2}} \ \zeta\left(2 (k+s),\left[ \frac{1}{\bar{\varepsilon}}\right] +1\right) \nonumber
\eeq
where $\zeta(k,a)$ is the Hurwitz zeta function.
Notice that $\Phi^{(2)}(s,\bar{\varepsilon})$ has an infinite number of singularities, located at semi-integer values $s = \frac{1}{2} -j$, with $j=0,1,\dots$

\begin{figure}
~\bigskip\bigskip
\begin{center}
\includegraphics[width=8cm]{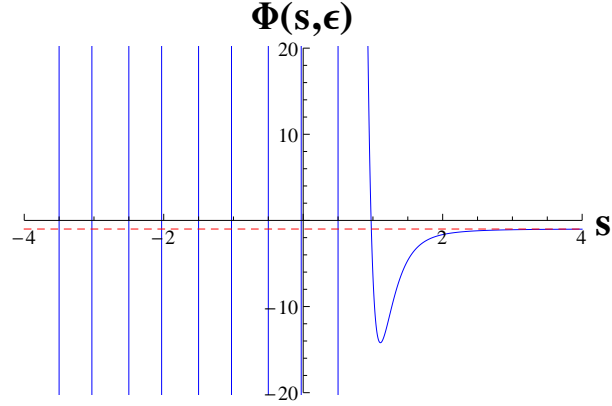}
\caption{$\Phi(s,\bar{\varepsilon})$ of eq.(\ref{phis}) as function of $s$ for $\bar{\varepsilon} = 2/101$; 
the horizontal line is $\lim_{s \rightarrow \infty}  \Phi^{(1)}(s,\bar{\varepsilon})$.
}
\bigskip
\label{Figure3}
\end{center}
\end{figure}

For $|\eta| \ll 1$ the spectral zeta of this string is 
\beq
Z(s) &\approx& 
\left( \frac{8 L^{2}}{\pi^2}\right)^s \zeta(2 s) + s \eta \frac{\bar{\varepsilon}^3   2^{3 s-1}  L^{2 s}}{\pi ^{2 s+1} } 
\sin\left(\frac{\pi}{\bar{\varepsilon}}\right) \sin\left(\frac{\pi  \ell}{\bar{\varepsilon}L}\right) 
 \Phi(s,\bar{\varepsilon}) \ .  \nonumber
\eeq

For $s\rightarrow -1/2$, this spectral zeta function to first order in $\eta$ reads
\beq
Z(s) &\approx& - \frac{\pi}{24 \sqrt{2} L} -\frac{\bar{\varepsilon}^3 \eta  
\sin\left(\frac{\pi}{\bar{\varepsilon}}\right)  \sin\left(\frac{\pi  \ell}{\bar{\varepsilon}L}\right)}{8 \sqrt{2} L} \left[\Phi^{(1)}(-1/2,\bar{\varepsilon}) + 
\sum_{k=0,k\neq 1}^\infty \frac{1}{\bar{\varepsilon}^{2k-2}} \ \zeta \left(2 k-1,\left[ \frac{1}{\bar{\varepsilon}}\right] +1\right) \right. \nonumber \\
&-& \left. \left(-H_{\left[ \frac{1}{\bar{\varepsilon}}\right] }+\frac{1}{2 s+1}+\gamma \right)\right] \nonumber \ ,
\eeq
where $H_{n} = \sum_{k=1}^n \frac{1}{k}$ is the harmonic number of order $n$. 
Notice that the term corresponding to $k=1$ in the series is treated separately because of the singularity of the Hurwitz zeta function.

It is also useful to write:
\beq
\Phi^{(1)}(-1/2,\bar{\varepsilon}) - H_{\left[ \frac{1}{\bar{\varepsilon}}\right] } = \sum_{n=1}^{[\frac{1}{\bar{\varepsilon}}]} \left[
\frac{n}{n^2\bar{\varepsilon}^2-1} - \frac{1}{n} \right] = \frac{H_{\left[ \frac{1}{\bar{\varepsilon}}\right] -\frac{1}{\bar{\varepsilon}
   }}}{2 \bar{\varepsilon} ^2}+\frac{H_{\left[ \frac{1}{\bar{\varepsilon}}\right]
   +\frac{1}{\bar{\varepsilon} }}}{2 \bar{\varepsilon} ^2}-\frac{H_{-\frac{1}{\bar{\varepsilon} }}}{2
   \bar{\varepsilon} ^2}-\frac{H_{\frac{1}{\bar{\varepsilon} }}}{2 \bar{\varepsilon} ^2}-H_{\left[ \frac{1}{\bar{\varepsilon}}\right] } \nonumber \ .
\eeq

Since $Z(-1/2)$ diverges for $n \neq 1/\bar{\varepsilon}$ or for $\ell/\bar{\varepsilon}L$ non integer, the Casimir energy of this string is infinite in these cases. Moreover, in this example, there is an infinite sequence of singularities of the spectral zeta function located at negative semi-integer values of $s$.

It is easy to see that this divergence is absent if the original density is replaced by a piecewise constant curve, for instance $\tilde{\Sigma}(x) =  \Sigma( \frac{x_k+x_{k+1}}{2})$ for $x_k \leq x \leq x_{k+1}$. In this case the diagonal matrix  elements of $\tilde{\Sigma}(x)$ do not contain the $1/n^2$ term and the resulting spectral functions are finite at $s = -1/2$. 
Calling $h \equiv x_{k+1}- x_k \equiv {2L/N}$, this result holds for arbitrary small (but finite) $h$ (or equivalently for arbitrary large $N$). Here $h$ acts as a cutoff of the highly excited states with quantum numbers  $n \gg 1/h$.

Let us discuss in detail this approach; in this case we have
\beq
\sigma(L) = 2\sqrt{2} L + \frac{L \eta}{\sqrt{2} N}  \sin \left(\frac{\pi }{\bar{\varepsilon}}\right) \csc \left(\frac{\pi }{\bar{\varepsilon} N}\right) \sin \left(\frac{\pi  \ell}{\bar{\varepsilon} L}\right) + O\left(\eta^2 \right) \nonumber
\eeq
and $\bar{\Sigma} = \sigma(L)^2/4L^2$.

The diagonal matrix elements of $\tilde{\Sigma}$ may be evaluated explicitly and read:
\beq
\langle n | \tilde{\Sigma} | n \rangle =    2 + \frac{\eta  \sin \left(\frac{\pi }{\bar{\varepsilon}}\right) \sin
   \left(\frac{\pi  \ell}{\bar{\varepsilon} L}\right) \left(2 \pi  n
   \csc \left(\frac{\pi }{\bar{\varepsilon} N}\right)-N
   \sin \left(\frac{\pi  n}{N}\right) \left(\csc
   \left(\frac{\pi -\pi  \bar{\varepsilon} n}{\bar{\varepsilon}
   N}\right)+\csc \left(\frac{\pi  \bar{\varepsilon} n+\pi
   }{\bar{\varepsilon} N}\right)\right)\right)}{2 \pi  n N} \ . \nonumber
\eeq

Working to first order in $\eta$ we have
\beq
Z(s) 
&\approx& 2^{3 s-1} \pi ^{-2 s} L^{2 s}  \left(2+\frac{\eta}{N}  s 
\sin\left(\frac{\pi }{\bar{\varepsilon}}\right) \csc \left(\frac{\pi}{\bar{\varepsilon} N}\right) 
\sin \left(\frac{\pi \ell}{\bar{\varepsilon} L}\right)\right) \zeta (2 s) \nonumber \\
&+& \eta  \left(-2^{3 s-2}\right) \pi ^{-2 s-1} s \sin \left(\frac{\pi}{\bar{\varepsilon}}\right) L^{2 s} 
\sin \left(\frac{\pi \ell}{\bar{\varepsilon} L}\right) 
\sum_{n=1}^\infty n^{-2 s-1} \ \sin \left(\frac{\pi n}{N}\right) 
\left(\csc \left(\frac{\pi -\pi \bar{\varepsilon} n}{\bar{\varepsilon} N}\right)
+\csc\left(\frac{\pi  \bar{\varepsilon} n+\pi }{\bar{\varepsilon} N}\right)\right) \ .  \nonumber
\eeq

The function $W(n) \equiv\sin \left(\frac{\pi n}{N}\right) 
\left(\csc \left(\frac{\pi -\pi \bar{\varepsilon} n}{\bar{\varepsilon} N}\right)
+\csc\left(\frac{\pi  \bar{\varepsilon} n+\pi }{\bar{\varepsilon} N}\right)\right)$ is periodic with period $N$, 
$W(n+N) = W(n)$, and therefore we may write
\beq
Z(s) 
&\approx&  2^{3 s-1} \pi ^{-2 s} L^{2 s}  \left(2+\frac{\eta}{N}  s 
\sin\left(\frac{\pi }{\bar{\varepsilon}}\right) \csc \left(\frac{\pi}{\bar{\varepsilon} N}\right) 
\sin \left(\frac{\pi \ell}{\bar{\varepsilon} L}\right)\right) \zeta (2 s) \nonumber \\
&+& \eta  \left(-2^{3 s-2}\right) \pi ^{-2 s-1} s \sin \left(\frac{\pi}{\bar{\varepsilon}}\right) L^{2 s} 
\sin \left(\frac{\pi \ell}{\bar{\varepsilon} L}\right) N^{-2 s-1} \nonumber\\
&\cdot& \sum_{n=1}^{N-1}  \ \sin \left(\frac{\pi n}{N}\right) 
\left(\csc \left(\frac{\pi -\pi \bar{\varepsilon} n}{\bar{\varepsilon} N}\right)
+\csc\left(\frac{\pi  \bar{\varepsilon} n+\pi }{\bar{\varepsilon} N}\right)\right) \ 
\zeta \left(2 s+1,\frac{n}{N}\right) \ .  \nonumber
\eeq

For a finite $N$, this expression is finite at $s=-1/2$: the divergent behavior of the Casimir energy of the string with a smooth
density is recovered only in the limit $N \rightarrow \infty$, where the sum over $n$ diverges.

As a further example, we may calculate the spectral sum rule at $s=1$ for a string with an arbitrary density of 
period $\Delta$ and length $2L=1$:
\beq
\Sigma(x) = \frac{a_0}{2} + \sum_{j=1}^\infty \left[ a_j \ \cos \frac{2\pi j x}{\Delta} + 
b_n \ \sin \frac{2\pi j x}{\Delta} \right] \nonumber   \ ,
\eeq
where 
\beq
a_0 = \frac{M}{L} - \sum_{j=1}^\infty \frac{a_j \Delta}{j L \pi} \ \sin \frac{2 L j \pi}{\Delta} \nonumber 
\eeq
and $M \equiv \int_{-L}^{L} \Sigma(x) dx$ is the total mass of the string.

The diagonal matrix elements of the density in this case are
\beq
\langle n | \Sigma | n \rangle &=&  \frac{a_0}{2} - \sum_{j=1}^\infty \frac{\Delta^3 n^2 a_j 
\sin \left(\frac{2 \pi  j L}{\Delta}\right)}{8 \pi j^3 L^3-2 \pi  \Delta^2 j L n^2} \nonumber
\eeq
and   
\beq
Z(1) &=& \sum_{n=1}^\infty \frac{\langle n | \Sigma | n \rangle}{n^2\pi^2/4L^2} 
=  \frac{M L}{3} + \sum_{j=1}^\infty 
\left[\frac{\Delta^3 a_j \sin \left(\frac{2 \pi  j L}{\Delta}\right)}{4 \pi ^3 j^3
   L}-\frac{\Delta^2 a_j \cos \left(\frac{2 \pi  j L}{\Delta}\right)}{2 \pi
   ^2 j^2} 
   -\frac{\Delta L a_j \sin \left(\frac{2 \pi  j L}{\Delta}\right)}{3
   \pi  j}\right]  \ . \nonumber
\eeq

Notice that $Z(1)$ reduces to the sum rule of a homogeneous string for $\Delta = 2L/N$ with $N$ integer.

\subsection{Inhomogeneous drums}

In this section we will discuss the calculation of the spectral zeta functions of inhomogeneous drums.

\subsubsection{Spectral zeta function of a square membrane}

We consider a square of side $2L$: in this case the eigenvalues of the negative laplacian 
(with Dirichlet boundary conditions at the border) are known exactly and read
\beq
E_{n_x,n_y} = \frac{\pi^2}{4L^2} (n_x^2+n_y^2) \nonumber \ ,
\eeq
with $n_x, n_y = 1,2, \dots$.

The series defining the spectral zeta function for this problem,
\beq
Z_{\Box}(s) = \sum_{n_x,n_y=1}^\infty \frac{1}{E_{n_x,n_y}^s} \nonumber \ ,
\eeq
may be analytically continued to all $s$. Ziff~\cite{Ziff86} and Steiner~\cite{Steiner87} report the 
spectral zeta function of the square: using the notation of Ziff it reads 
\beq
Z_{\Box}(s) = \left(\frac{2L}{\pi}\right)^{2s}  (\beta (s) \zeta (s)-\zeta (2 s)) \nonumber \ ,
\eeq
where $\beta(s) \equiv \sum_{n=0}^\infty (-1)^n (2n+1)^{-s}$. 
Notice that $s=1/2$ and $s=1$ are the only singularities of $Z_\Box(s)$.

Around $s=1$ the zeta function behaves as
\beq
Z_{\Box}(s) &\approx& \frac{\mathcal{A}_\Box}{4 \pi} \ \left\{ \frac{1}{s-1} + 
\left(\log \left(\frac{\mathcal{A}_\Box}{4 \pi ^2}\right)+\frac{\gamma _1\left(\frac{3}{4}\right)-\gamma
   _1\left(\frac{1}{4}\right)}{\pi }-\frac{2 \pi }{3}+\gamma \right) +
\dots \right\} \nonumber  \\
&\equiv& \frac{\mathcal{A}_\Box}{4 \pi} \ \left\{ \frac{1}{s-1} + g_{\Box} + \dots \right\} \nonumber \ , 
\eeq
where $\mathcal{A}_\Box \equiv 4L^2$ is the area of the square. $\gamma \approx 0.577216$ is the Euler--Mascheroni gamma and
$\gamma_1(a)$ is the Stieltjes constant defined as  $\gamma_1(a) = \lim_{s\rightarrow 1} \left( \frac{1}{s-1} -\zeta(s,a)\right)$ 
($\zeta(s,a) = \sum_{n=0}^\infty \frac{1}{(n+a)^s}$ is the Hurwitz zeta function).

\subsubsection{Spectral zeta function on an arbitrary simply connected domain in two dimensions}

Let $\mathcal{D}$ be an arbitrary simply connected domain of the plane  and call $E_n$ the eigenvalues of the Helmholtz equation on this domain, assuming Dirichlet boundary conditions on $\partial \mathcal{D}$. Riemann's theorem grants the existence of a conformal map from $\mathcal{D}$ to any other simply connected region of the plane and in particular to $(x,y) \in \Omega$, $\Omega = \left\{ |x| \leq L, |y|\leq L\right\}$:
\beq
w = u + i v = f(z) \ , \nonumber
\eeq
$(u,v) \in \mathcal{D}$.

Under this map, the original Helmholtz equation on $\mathcal{D}$ transforms into the Helmholtz equation on $\Omega$, in presence of a variable density $\Sigma(x,y) = \left| \frac{df}{dz} \right|^2$ (see ref.~\cite{Amore10}). We will refer to $\Sigma(x,y)$ as a {\sl conformal
density}, to underline that it does not correspond to a physical density of the membrane.

We observe that, if $\Sigma(x,y)$ is a conformal density, its integral over the square is just the area of the region $\mathcal{D}$, $\int_{-L}^{+L}dx \int_{-L}^{+L} dy \Sigma(x,y) = \mathcal{A}$. 

For $s \rightarrow 1^+$ we may obtain a general behavior of the spectral zeta function to order $O\left(s-1\right)$ as
\beq
\left. Z(s) \right|_{s\rightarrow 1^+} &\approx& \sum_{n=1}^\infty \left( \frac{\langle n | \Sigma| n \rangle}{\epsilon_n} \right)^s
+ O\left(s-1\right) \nonumber \\
&\approx& \sum_{n=1}^\infty \left( \frac{\mathcal{A} }{4 L^2 \epsilon_n} \right)^s 
+ \sum_{n=1}^\infty \left[ \left( \frac{\langle n | \Sigma| n \rangle}{\epsilon_n} \right)^s - 
\left( \frac{\mathcal{A} }{4 L^2 \epsilon_n} \right)^s\right] +
O\left(s-1\right) \nonumber  \ , 
\eeq
where the second series converges for $s = 1$. The first series, on the other hand, reproduces the spectral zeta function
of a square of side $2L$, apart for a multiplicative factor $\left( \frac{\mathcal{A} }{4 L^2} \right)^s$.
For this reason, we may obtain the general behavior of the spectral zeta function at $s=1$ as:
\beq
\left. Z(s) \right|_{s\rightarrow 1^+} &\approx&  \frac{\mathcal{A}}{4 \pi} \ \left\{ \frac{1}{s-1} + 
g + \dots \right\} \nonumber  \ , 
\eeq
where
\beq
g &\equiv& g_{\Box} + \log \frac{\mathcal{A}}{4L^2} + \frac{4\pi}{\mathcal{A}} \ \sum_{n=1}^\infty \left[ \frac{\langle n | \Sigma| n \rangle - \frac{\mathcal{A} }{4 L^2} }{\epsilon_n}\right]  \nonumber \\
&=& \left(\log \left(\frac{\mathcal{A}}{4 \pi ^2}\right)+\frac{\gamma _1\left(\frac{3}{4}\right)-\gamma
   _1\left(\frac{1}{4}\right)}{\pi }-\frac{2 \pi }{3}+\gamma \right) + \frac{4\pi}{\mathcal{A}} \ \sum_{n=1}^\infty \left[ \frac{\langle n | \Sigma| n \rangle - \frac{\mathcal{A} }{4 L^2} }{\epsilon_n}\right]  \nonumber  \ .
\eeq

Itzykson, Moussa  and Luck have found this behavior long time ago ( eq.(6) of  ref.~\cite{Itzykson86}), obtaining an expression for
$g$ in terms of an integral.

\subsubsection{Spectral zeta function of a deformed squared membrane }

\begin{figure}
~\bigskip\bigskip
\begin{center}
\includegraphics[width=4cm]{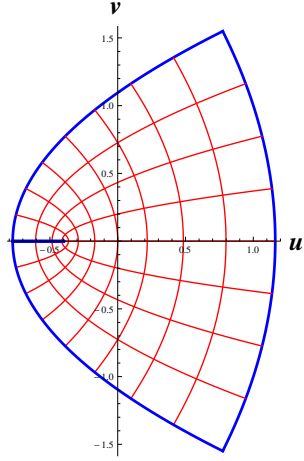}
\caption{Deformation of the square for $\alpha=1/2$.}
\bigskip
\label{Figure4}
\end{center}
\end{figure}

As a specific example, we consider the domain obtained from the map
\beq
f(z) = \frac{z + \alpha z^2}{\sqrt{1+\frac{8 \alpha ^2 L^2}{3}}} \nonumber
\eeq
for $|\alpha| < 1/2$ (we have studied this problem for $|\alpha|\ll 1$ in a previous paper, ref.\cite{Amore10}, 
calculating the lowest energies of this domain in perturbation theory and comparing them with the 
numerical values obtained using a collocation approach). The domain corresponding to $\alpha =1/2$ 
is shown in Fig.\ref{Figure4}.

The map is defined to allow $\mathcal{A} = \mathcal{A}_\Box$; the conformal density in this case is 
\beq
\Sigma(x,y) = \frac{3 \left(4 \alpha ^2 y^2+(2 \alpha  x+1)^2\right)}{8 \alpha ^2 L^2+3} \nonumber
\eeq
and its matrix elements in the basis of the square are
\beq
\langle n_x, n_y | \Sigma | n'_x, n'_y \rangle &=& \left\{
\begin{array}{lll}
1 -\frac{6 \mathcal{A} \alpha ^2 \left(n_x^2+n_y^2\right)}{\pi ^2 n_x^2 n_y^2 \left(2 \mathcal{A} \alpha ^2+3\right)} & , & 
n_x = n'_x \ \ ; \ \ n_y = n'_y \\
\frac{48 \mathcal{A} \alpha ^2 n_y n'_y \left((-1)^{n_y+n'_y}+1\right)}{\pi ^2 \left(2 \mathcal{A} \alpha ^2+3\right) (n_y-n'_y)^2
   (n_y+n'_y)^2} & , & n_x = n'_x \ \ ; \  \ n_y \neq n'_y \\
\frac{48 \sqrt{\mathcal{A}} \alpha  n_x n'_x \left(\sqrt{\mathcal{A}} \alpha  \left((-1)^{n_x+n'_x}+1\right)+(-1)^{n_x+n'_x}-1\right)}{\pi ^2
   \left(2 \mathcal{A} \alpha ^2+3\right) (n_x-n'_x)^2 (n_x+n'_x)^2}  & , & 
n_x \neq n'_x \ \ ; \ \  n_y = n'_y \\   
0 & , & 
n_x \neq n'_x \ \ ; \ \ n_y \neq n'_y \\
\end{array} 
\right. \nonumber
\eeq

Using the diagonal matrix elements above, one finds in this case
\beq
g 
&=& \left(\log \left(\frac{\mathcal{A}}{4 \pi ^2}\right)+\frac{\gamma _1\left(\frac{3}{4}\right)-
    \gamma_1\left(\frac{1}{4}\right)}{\pi }-\frac{2 \pi }{3}+\gamma \right)  
- \frac{2 \pi \mathcal{A} \alpha^2}{3 \left(2 \mathcal{A} \alpha ^2+3\right) }  \nonumber \ .
\eeq

In Table \ref{table-1} we report the values of the spectral zeta function at $s=2$ for the deformed square for different values of $\alpha$:
the exact value,  $Z(2)$, is compared with the value obtained considering only the diagonal contributions,  $Z(2)^{(diag)}$,
and with the  values obtained from the first $3000$ numerical eigenvalues obtained with a collocation approach with a grid with $9801$ points 
(adding the contribution of the higher excited states via the Weyl's law):
\beq
Z(2)^{(num)} \equiv \sum_{n=1}^{3000} \left(\frac{1}{E_n^{(num)}} \right)^2 + 
\sum_{n=3001}^\infty \left(\frac{1}{\frac{4\pi n}{\mathcal{A}} + \frac{\mathcal{L}}{\mathcal{A}}\sqrt{\frac{4\pi n}{\mathcal{A}}}}\right)^2
\nonumber \ ,
\eeq
where $\mathcal{A}=4$ is the area of the domain and $\mathcal{L}$ its perimeter; in the last column we report the value obtained only using Weyl's law:
\beq
Z(2)^{(Weyl)} \equiv 
\sum_{n=1}^\infty \left(\frac{1}{\frac{4\pi n}{\mathcal{A}} + \frac{\mathcal{L}}{\mathcal{A}}\sqrt{\frac{4\pi n}{\mathcal{A}}}}\right)^2 \nonumber \ .
\eeq

Notice that $Z(2)^{(diag)}$ accounts almost completely for the value of $Z(2)$, up to $\alpha \approx 1/10$.

\begin{table}[!htb]
\begin{tabular}{|c||c|c|c|c|}
	\hline
$\alpha$	    & $Z(2)$ &   $Z(2)^{(diag)}$ & $Z(2)^{(num)}$ & $Z(2)^{(Weyl)}$   \\
	\hline \hline
$\frac{1}{100}$ & 0.06970508 &  0.06968939  &  0.06970508 & 0.04950760\\
$\frac{1}{50}$  & 0.06969987 &  0.06963720  &  0.06969987 & 0.04950758\\
$\frac{1}{25}$  & 0.06967869 &  0.06942953  &  0.06967869 & 0.04950735\\
$\frac{1}{10}$  & 0.06951485 &  0.06802167  &  0.06951486 & 0.04949801\\
$\frac{1}{4}$   & 0.06805735 &  0.06073539  &  0.06805740 & 0.04918833\\
$\frac{1}{2}$   & 0.06143122 &  0.04641541  &  0.06143131 & 0.04662053\\
	\hline \hline
\end{tabular}
\caption{Spectral zeta function at $s=2$ for the deformed square of area $\mathcal{A}=4$ for different values of $\alpha$. }
\label{table-1}
\end{table}

\subsubsection{Spectral zeta function of the annulus}

The function
\beq
f(z) &=& e^{z-L}   \nonumber \ , 
\eeq
maps the rectangle $[-L,L] \times [-\pi, \pi]$ onto a circular annulus of external radius $R=1$ and internal radius $r = e^{-2L}$.
The conformal density in this case is 
\beq
\Sigma(x,y) = e^{2 (x-L)} \nonumber  \ , 
\eeq
and depends only on $x$.

Since we are interested in calculating the Casimir energy of this configuration we need to evaluate the contributions of both the transverse electric (TE) and transverse magnetic (TM) modes. The first ones correspond to imposing Dirichlet boundary conditions at the border of the annulus, while the second ones correspond to imposing Neumann boundary conditions at the border. 

In the case of TE modes we use the basis on the rectangle is (see ref.\cite{Alvarado})
\beq
\Psi_{n_x,n_y,u}(x,y) = \psi_{n_x}(x) \times \left\{ \begin{array}{ccc}
\chi_{n_y}(y) & , & u=1\\
\phi_{n_y}(y) & , & u=2\\
\end{array}
\right.  \nonumber
\eeq
where
\beq
\psi_{n_x}(x) = \frac{1}{L} \sin \left(\frac{n_x\pi}{2L} (x+L) \right) \ \ , \ \ n_x = 1,2, \dots \nonumber
\eeq
and
\beq
\chi_{n_y}(y) &=& \left\{ \begin{array}{ccc}
\frac{1}{\sqrt{2\pi}} & , & n_y= 0\\
\frac{1}{\sqrt{\pi}} \sin (n_y y) &, &  n_y=1,2,\dots \\
\end{array}
\right.  \\
\phi_{n_y}(y) &=& \frac{1}{\sqrt{\pi}} \cos (n_y y)  \ \ , \ \ n_y = 1,2,\dots  \ , \nonumber
\eeq
which corresponds to Dirichlet boundary conditions at $x= \pm L$ and periodic boundary conditions
at $y=\pm \pi$. Each state is determined by three quantum numbers, $n_x$, $n_y$ and $u$.
Alternatively we may also impose Dirichlet boundary condition on both directions: this case corresponds to a circular annulus with a cut. 

The eigenvalues of the negative laplacian on this domain are
\beq
\epsilon_{n_x,n_y,u}^{(DP)} = \frac{n_x^2\pi^2}{4L^2} + n_y^2 \ , \  u=1,2 \nonumber \ ,
\eeq
and
\beq
\epsilon_{n_x,n_y}^{({\rm DD})} = \frac{n_x^2\pi^2}{4L^2} + \frac{n_y^2}{4} \nonumber \ ,
\eeq
where the superscripts (DP) and (DD) refer to Dirichlet-Periodic and Dirichlet-Dirichlet boundary conditions
respectively.  We appreciate that the states DP with $n_y \geq 1$ are doubly degenerate, whereas states with $n_y=0$ are
non degenerate.

The matrix elements of $\Sigma$ are the same in both basis since the density does not depend on $y$:
\beq
\langle n_x, n_y, u | \Sigma | n'_x , n'_y , u' \rangle^{(DP)} &=& \langle n_x, n_y| \Sigma | n'_x , n'_y  \rangle^{({\rm DD})} \nonumber \\
&=&  \left\{ 
\begin{array}{ccc}
\frac{\pi ^2 n_x^2 \left(r^2-1\right)}{2 \left(\pi ^2 n_x^2
   \log (r)+\log ^3(r)\right)} & , & n_x = n'_x , n_y = n'_y , u=u'\\
-\frac{8 \pi ^2 n_x n'_x \log (r)
   \left((-1)^{n_x+n'_x}-r^2\right)}{\left(\pi ^2
   (n_x-n'_x)^2+4 \log ^2(r)\right) \left(\pi ^2
   (n_x+n'_x)^2+4 \log ^2(r)\right)} & , & n_x \neq n'_x , n_y = n'_y , u=u'\\
\end{array}
\right. \  . \nonumber
\eeq
Notice that the (DD) matrix elements only hold for $n_y \geq 1$.

\begin{figure}
~\bigskip\bigskip
\begin{center}
\includegraphics[width=6cm]{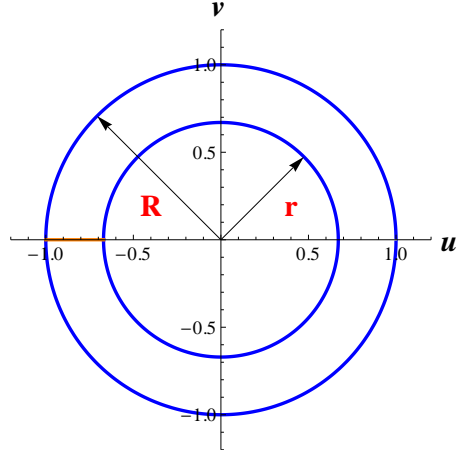}
\caption{Circular annulus.}
\bigskip
\label{fig_annulus}
\end{center}
\end{figure}

We will first concentrate on the circular annulus (DP): the value of its spectral zeta function  $s=2$ is 
\beq
Z^{(DP)}(2) = \sum_{n_x,n_y,u} \ \sum_{n'_x} \frac{\langle n_x, n_y, u | \Sigma | n'_x , n_y , u\rangle^2}{\epsilon_{n_x,n_y,u}^{(DP)}\epsilon_{n'_x,n_y,u}^{(DP)}} \ .
 \nonumber
 \eeq

In the case of the circular annulus the exact eigenfunctions of the negative laplacian are known and read
\beq
\Phi_{mns}(\rho,\theta) = N_{mns} \left[ Y_m(k_{mn}) J_m(k_{mn}\rho) - J_m(k_{mn}) Y_m(k_{mn}\rho)  \right] \times \left\{ \begin{array}{ccc}
\cos n\theta & , & u=1\\
\sin n\theta & , & u=2\\
\end{array}
\right. \nonumber \ , 
\eeq
where $r \leq \rho \leq 1$; $N_{mns}$ is a normalization constant and $J$ and $Y$ are the Bessel functions of first and second kind. 
The corresponding eigenvalues are obtained from the solutions of the equation
\beq
Y_m(k) J_m(kr) - J_m(k) Y(kr) = 0  \ ,
\label{egv}
\eeq
as
\beq
E_{mn} = k_{mn}^2 \nonumber \ .
\eeq

In ref.\cite{Alvarado} we have observed that the energies of thin annuli are approximated 
extremely well by the formula
\beq
E_{n_x,n_y,s} \approx \frac{\epsilon^{(DP)}_{n_x,n_y,u}}{\langle n_x, n_y, u | \Sigma | n_x , n_y , u \rangle} \nonumber \ .
\eeq
This formula not only contains the correct asymptotic behavior (Weyl's law) but it also describes
quite accurately the low energy behavior of the eigenvalues (see for instance Fig.2 of ref.\cite{Alvarado}).
In this case one expects that most of the contribution to $Z(2)$ must be carried by the diagonal terms:
\beq
Z^{(diag)}(2) = \sum_{n_x,n_y,u} \frac{\langle n_x, n_y, u | \Sigma | n_x , n_y , u\rangle^2}{(\epsilon^{(DP)}_{n_x,n_y,u})^2} \ .
\eeq

In Table \ref{table-2} we compare the exact value $Z^{(DP)}(2)$, with $Z^{(diag)}(2)$ and with 
\beq
Z^{(DP)}(2)^{(num)} &\equiv& \sum_{n=1}^{10000} \left(\frac{1}{E_n^{(num)}} \right)^2 + 
\sum_{n=10001}^\infty \left(\frac{1}{\frac{4\pi n}{\mathcal{A}} + \frac{\mathcal{L}}{\mathcal{A}}\sqrt{\frac{4\pi n}{\mathcal{A}}}}\right)^2 \nonumber \\
Z^{(DP)}(2)^{(Weyl)} &\equiv& 
\sum_{n=1}^\infty \left(\frac{1}{\frac{4\pi n}{\mathcal{A}} + \frac{\mathcal{L}}{\mathcal{A}}\sqrt{\frac{4\pi n}{\mathcal{A}}}}\right)^2 \nonumber \ ,
\eeq
where $\mathcal{A}=\pi (1-r^2)$ is the area of the circular annulus and $\mathcal{L} = 2 \pi (1+r)$ its perimeter.
$E_n^{(num)}$ are the first $10000$ numerical eigenvalues obtained solving the trascendental equation (\ref{egv}), each calculated
with a precision of $20$ digits.

\begin{table}[!htb]
\begin{tabular}{|c||c|c|c|c|}
	\hline
$r$	    & $Z(2)$ &   $Z(2)^{(diag)}$ & $Z(2)^{(num)}$ & $Z(2)^{(Weyl)}$   \\
	\hline \hline
$1/10$ & 0.0257710759 &  0.0169570674 &  0.0257710743 & 0.030450016 \\
$1/2$  & 0.0057419570 &  0.0054705758 &  0.0057419569 & 0.011541075 \\
$9/10$ & 0.0000578599 &  0.0000577934 &  0.0000578601 & 0.000251957 \\
         \hline \hline
\end{tabular}
\caption{Spectral zeta function at $s=2$ for a circular annulus of varying internal radius $r$ and
fixed external radius $R=1$. }
\label{table-2}
\end{table}

Notice that for $r=9/10$ the off-diagonal contributions account only for about $1/1000$ of the value of $Z(2)$, confirming the observation made in ref.\cite{Alvarado}; the value obtained using Weyl's law on the other hand largely overestimates $Z(2)$, since this approximation fails to describe the low energy part of the spectrum.

It is interesting to obtain the quantity
\beq
\Delta Z(1)  \equiv \lim_{s\rightarrow 1^+} \left[Z^{(DP)}(s) -Z^{({\rm DD})}(s) \right] \nonumber \ ,
\eeq
which is finite.

We have
\beq
\Delta Z(1) &=& \sum_{n_x=1}^\infty \frac{\langle n_x, 0, 1 | \Sigma | n_x , 0 , 1\rangle^2}{\left(\frac{n_x \pi}{2L}\right)^2} + 
\sum_{n_x=1}^\infty \ \sum_{n_y=1}^\infty \langle n_x, ny | \Sigma | n_x , ny \rangle^2 \ 
\left[ \frac{2}{\frac{n_x^2\pi^2}{4L^2}+n_y^2} - \frac{1}{\frac{n_x^2\pi^2}{4L^2}+\frac{n_y^2}{4}}
\right] \nonumber \\
&=& \frac{\left(1-e^{-4 L} \right) (2 L \coth (2 L)-1)}{16 L}
+\sum_{n_x=1}^\infty \frac{\pi ^2 e^{-2 L} n_x \sinh (2 L) \text{csch}\left(\frac{\pi^2 n_x}{L}\right)}{4 L^2+\pi ^2 n_x^2}  \nonumber \\
&=& \frac{1}{8 \log (r)} \left[ (1-r^2)+\left(1+r^2\right) \log (r)\right] \nonumber \\
&+& \frac{1}{2} \left(1-r^2\right) \sum_{j=0}^\infty
\left[ B\left( {e^{\frac{2 (2 j+1) \pi^2}{\log (r)}}}, 1-\frac{i \log (r)}{\pi},0\right)
+B\left( e^{\frac{2 (2 j+1) \pi ^2}{\log (r)}},\frac{i\log (r)}{\pi }+1,0\right)\right]\nonumber \ ,
\eeq
where $B(z,a,b)$ is the incomplete beta function. This expression is exact and holds for all values $r$, $0 < r < 1$: in particular, around 
$r=1$ we have
\beq
\Delta Z(1) &\approx& \frac{1}{12} (1-r)^2 \nonumber  \ .
\eeq

We will now discuss the analytic continuation of the spectral zeta function of the annulus, in the limit $r \rightarrow 1^-$ ($L \rightarrow 0^+$).
In this limit, it is possible to approximate the conformal density as
\begin{eqnarray}
\Sigma(x,y) \approx (1-2L) (1+2 x)  + \dots \nonumber
\end{eqnarray}
and
\begin{eqnarray}
\delta\Sigma(x,y) \approx -2L + 2x (1-2L) + \dots \ . \nonumber
\end{eqnarray}

Working to this order we have that 
\begin{eqnarray}
\langle n_x,n_y, u | \delta\Sigma | n_x , n_y , u\rangle &=& -2L \nonumber
\end{eqnarray}
and the spectral zeta function of thin annulus may be approximated as
\begin{eqnarray}
Z^{(DP)}(s) \approx \sum_{n_x,n_y,u} \frac{1-2 s L}{\left(\epsilon^{(DP)}_{n_x,n_y}\right)^s} 
= \sum_{n_x=1}^\infty \frac{1-2 s L}{\left(\epsilon^{(DP)}_{n_x,0}\right)^s} + 
 2 \sum_{n_x=1}^\infty \sum_{n_y=1}^\infty \frac{1-2 s L}{\left(\epsilon^{(DP)}_{n_x,n_y}\right)^s}  \nonumber \ .
\end{eqnarray}

An analogous expression can be obtained for the case of a thin annulus with a trasversal cut, which corresponds to choosing Dirichlet boundary conditions along the two directions. In this case we have
\beq
\epsilon_{n_x,n_y}^{({\rm DD})} = \frac{n_x^2\pi^2}{4L^2} + \frac{n_y^2}{4} \nonumber \ ,
\eeq
where the superscript (DD) indicates Dirichlet-Dirichlet boundary condintions, and $n_x, n_y = 1,2, \dots$.
The spectral zeta function in this case reads
\begin{eqnarray}
Z^{({\rm DD})}(s) \approx \sum_{n_x,n_y} \frac{1-2 s L}{\left(\epsilon^{({\rm DD})}_{n_x,n_y}\right)^s} 
=  \sum_{n_x=1}^\infty \sum_{n_y=1}^\infty \frac{1-2 s L}{\left(\epsilon^{({\rm DD})}_{n_x,n_y}\right)^s} \ . \nonumber
\end{eqnarray}

The analytic continuation of these expressions is discussed explicitly by Kirsten in ref.~\cite{Kirsten10}:
following \cite{Kirsten10} we define
\beq
\zeta_{\mathcal{C}}(L_2,L_3,s) \equiv \sum_{\ell_2=1}^\infty \sum_{\ell_3=1}^\infty \left( \left(\frac{\pi \ell_2}{L_2}\right)^2
+\left(\frac{\pi \ell_3}{L_3}\right)^2\right)^{-s} \ . \nonumber
\eeq
The analytic continuation of this function is given in eq.(4.31) of \cite{Kirsten10}:
\beq
\zeta_{\mathcal{C}}(L_2,L_3,s) &=& - \frac{1}{2} \left(\frac{L_3}{\pi}\right)^{2s}  \zeta(2s)
+ \frac{L_2 \Gamma(s-1/2)}{2\sqrt{\pi} \Gamma(s)} \left(\frac{L_3}{\pi}\right)^{2s-1}  \zeta(2s-1) \nonumber \\
&+& \frac{2L_2^{s+1/2}}{\sqrt{\pi} \Gamma(s)} \sum_{\ell_2=1}^\infty \sum_{\ell_3=1}^\infty \left(\frac{\ell_2 L_3}{\pi \ell_3}\right)^{s-1/2} 
K_{1/2-s} \left(\frac{2\pi L_2 \ell_2\ell_3}{L_3}\right) \ . \nonumber
\eeq

Therefore
\begin{eqnarray}
Z^{(DP)}(s) &\approx& (1-2sL) \left[ 4^s \pi ^{-2 s} L^{2 s} \zeta (2 s) + 2 \ \zeta_{\mathcal{C}}(\pi,2L,s)  \right] \nonumber \\
Z^{({\rm DD})}(s) &\approx& (1-2sL) \ \zeta_{\mathcal{C}}(2\pi,2L,s)  \nonumber \ .
\end{eqnarray}

Around $s=-1/2$, in the limit $r \rightarrow 1^-$, we have
\beq
Z^{(DP)}(-1/2) \approx Z^{({\rm DD})}(-1/2) \approx -\frac{\zeta (3)}{4 (r-1)^2} + \dots \ ,
\eeq
which provides the transverse electric contribution to the Casimir energy of an arbitrarily thin annulus:
\beq
E_C^{(TE)} = \frac{1}{2} Z^{(DP)}(-1/2) \approx -\frac{\zeta (3)}{8 (r-1)^2} + \dots
\eeq

It is easy to calculate the transverse magnetic contribution to the Casimir energy keeping in mind 
that $\epsilon_{n,1}^{({\rm NN})} = \epsilon_{2n}^{({\rm DD})}$ and $\epsilon_{n,2}^{({\rm NN})} = \epsilon_{2n-1}^{({\rm DD})}$.
\beq
E_C^{(TM)} = \frac{1}{2} Z^{(NP)}(-1/2) \approx -\frac{\zeta (3)}{8 (r-1)^2} + \dots
\eeq

Thus the electromagnetic Casimir energy of a thin annulus is
\beq
E_C \approx -\frac{\zeta (3)}{4 (r-1)^2} + \dots
\eeq

\subsubsection{Spectral zeta function of concentric cylinders}

Gosdzinsky and Romeo\cite{Romeo98} have related the spectral zeta function of a circle to the spectral zeta function of a cylinder of
circular section; Nesterenko and Pirozhenko~\cite{Nesterenko00} have also used this formula to obtain the Casimir energy of a circle 
using the results for a cylinder. 

Using our conventions, the relation between the two spectral zeta functions, eq.(3) of \cite{Romeo98} and eq.(6) of \cite{Nesterenko00}) can
be cast in the form
\beq
Z_{cyl}(s)  = \frac{1}{2\pi} \ B\left(\frac{1}{2}, s-\frac{1}{2}\right) \ Z_{circle}\left(s-\frac{1}{2}\right) \ . 
\label{cylinder}
\eeq

Although the authors of  refs.~\cite{Romeo98, Nesterenko00} have applied this equation to the cylinder of circular section, the formula
is general and it applies to cylinders of arbitrary section: therefore we can obtain the spectral zeta functions for these cases, 
working in two dimensions with a domain corresponding to the transversal section of the cylinder and then
using eq.(\ref{cylinder}).

We write eq.(\ref{cylinder}) in the general form as
\beq
Z_{cyl}(s)  = \frac{1}{2\pi} \ B\left(\frac{1}{2}, s-\frac{1}{2}\right) \ Z\left(s-\frac{1}{2}\right) \ . 
\label{cylinder2}
\eeq

We can apply eq.(\ref{cylinder2}) to calculate the Casimir energy of an infinite cylinder with annular section, in the limit where
the two radii are very close ($r \rightarrow R^-$, $R=1$). The spectral zeta function of the two dimensional domain has been already 
calculated to leading order in the previous example, and thus the calculation only requires the straightforward application of 
eq.(\ref{cylinder2}):
\beq
Z^{(DP)}_{cyl}\left(-\frac{1}{2}\right)  \approx Z^{({\rm DD})}_{cyl}\left(-\frac{1}{2}\right)  \approx - \frac{\pi^3}{360} \frac{1}{(1-r)^3} \ .
\eeq

The total electromagnetic Casimir energy will therefore be
\beq
E_C = \frac{1}{2} \left[ Z^{(DP)}_{cyl}\left(-\frac{1}{2}\right) + Z^{(NP)}_{cyl}\left(-\frac{1}{2}\right) \right]
=  - \frac{\pi^3}{360} \frac{1}{(1-r)^3}  \ .
\eeq

The Casimir energy for concentric cylinders has been calculated by Mazzitelli and collaborators in \cite{Mazzitelli03}: our result reproduces 
eq.(61) of that paper\footnote{Notice however that the calculation of \cite{Mazzitelli03} is not restricted to the particular limit that we are
studying, $r \rightarrow 1^-$.}. This provides a useful check of the correctness of our approach.

The same approach can also be applied to the calculation of the Casimir energy of cylinders and annular cylinders with slightly nonuniform density.

\section{Conclusions}
\label{conclusions}

In this paper we have discussed a novel approach to the calculation of the spectral zeta function 
and of the heat kernel associated with the eigenvalues of a slighlty inhomogeneous system in a 
$d$-dimensional cube within perturbation theory. 
The method that we have devised relies on the standard Rayleigh-Schr\"odinger perturbation theory for hermitian operators and provides an explicit expression for the spectral zeta function associated to these operators that converges for $s>d/2$. To the best of our knowledge, this is the first calculation where the spectral zeta function of a system with arbitrary density perturbation has been obtained.  

The analytic continuation of this perturbative series allows one to obtain an 
approximation to  the Casimir energy of the system under consideration. We have
illustrated our method with some examples, evaluating the Casimir energy
of different systems in one, two and three dimensions. In particular, for a string 
with piecewise constant density and of two perfectly conduting, concentric 
cylinders of similar radius we have reproduced results already published~\cite{Lambiase00, Mazzitelli03}. This provides a useful check of our approach.

An interesting outcome of our analysis is the appearance of irremovable divergences already to first order perturbation theory in the Casimir energy of one dimensional inhomogeneous systems, when either Dirichlet-Dirichlet or Neumann-Neumann boundary conditions are applied at the ends of the string, unless the density $\Sigma(x)$ of the string is such that $\Sigma'(-L) = \Sigma'(L)$. The physical interpretation of these singularities is not clear. On the other hand, we also have found that the sum of Dirichlet-Dirichlet and Neumann-Neumann zeta functions is free of divergences, 
to first order. 
   
For positive integer values of $s$, with $s>d/2$, our formalism provides exact (non-perturbative) 
sum rules that can be useful to test numerical and analytic approximations to the eigenvalues of the problem.

The calculations presented in this paper have been mostly performed to first order in perturbation theory: the analytic continuation of the higher order expressions, which involve multiple series, 
is a non trivial task for general densities (in the paper we have discussed a one dimensional problem where the second order calculation is easily obtained)~\footnote{We see some analogies
with the difficulties which are found in calculating the corrections to the Casimir forces beyond the 
"proximity force approximation" (PFA)~\cite{Derjaguin}. An approach to the calculation of these corrections has been recently put forward by Fosco and collaborators in ref.~\cite{Fosco}, and used later by Bimonte et al. in ref.~\cite{Jaffe}.}.  
We are currently studying a non-perturbative extension of the present approach that may allow to
describe a larger class of problems.

\begin{acknowledgments}
The author ackowledges support of Conacyt through the SNI fellowship.
\end{acknowledgments}

\end{document}